\documentclass[12pt,a4paper]{article}
\usepackage{graphicx}
\usepackage{amssymb,amsfonts,amsmath}
\usepackage{bm}
\usepackage{color}
\usepackage{cite}
\usepackage{fancyhdr}
\renewcommand{\headrulewidth}{0pt} 

\begin{document}

\newcommand{\EQ}{Eq.~}
\newcommand{\EQS}{Eqs.~}
\newcommand{\FIG}{Fig.~}
\newcommand{\FIGS}{Figs.~}
\newcommand{\TAB}{Tab.~}
\newcommand{\TABS}{Tabs.~}
\newcommand{\SEC}{Sec.~}
\newcommand{\SECS}{Secs.~}

\newcommand{\degC}{${^\circ}$C }
\newcommand{\degs}{${^\circ}$/s }

%%% --- remove before submission
\definecolor{COL_comment}{rgb}{ .1, .6, .1} 
\newcommand{\comment}[1]{{\color{COL_comment}{#1}}}
\newcommand{\old}[1]{{\color{blue}#1}}
\newcommand{\new}[1]{{\color{red}#1}}
\newcommand{\masuda}[1]{{\color{magenta}#1}}
\newcommand{\comr}[1]{\textcolor{red}{#1}}
\newcommand{\comb}[1]{\textcolor{blue}{#1}}
%%% ---

%%% title
\ \\
\ \\
\ \\
\noindent
{\LARGE Long-tail Behavior in Locomotion of \textit{Caenorhabditis}\\ \textit{elegans}%
%: Data Analysis and Computational Modeling
}
\ \\
\ \\

\large
%%% author
\noindent
Jun Ohkubo,${}^{a,1}$
Kazushi Yoshida,${}^{b}$
Yuichi Iino,${}^{b}$
and Naoki Masuda${}^{c,d,*}$
\ \\
\ \\

\footnotesize
\noindent
%${}^{a}$
%Institute for Solid State Physics, 
%${}^{b}$ 
%Department of Biophysics and Biochemistry, Graduate School of Science, 
%and ${}^{c}$
%Department of Mathematical Informatics,
%Graduate School of Information Science and Technology,
%The University of Tokyo, Tokyo, Japan;
%${}^{\P}$
%PRESTO, Japan Science and Technology Agency, Saitama, Japan
${}^{a}$
Institute for Solid State Physics,
The University of Tokyo,
5-1-5 Kashiwanoha, Kashiwa, Chiba 277-8581, Japan\\
%\ \\
${}^{b}$ 
Department of Biophysics and Biochemistry,
Graduate School of Science,
The University of Tokyo,
7-3-1 Hongo, Bunkyo, Tokyo 113-8656, Japan\\
%\ \\
${}^{c}$
Department of Mathematical Informatics,
Graduate School of Information Science and Technology,
The University of Tokyo,
7-3-1 Hongo, Bunkyo, Tokyo 113-8656, Japan\\
%\ \\
${}^{d}$
PRESTO, Japan Science and Technology Agency,
4-1-8 Honcho, Kawaguchi, Saitama 332-0012, Japan\\
%\ \\

\ \\
\ \\
\noindent
$^{*}$ Correspondence: masuda@mist.i.u-tokyo.ac.jp\\
$^{1}$ Present address: Graduate School of Informatics, Kyoto University, 
36-1 Yoshida-Honmachi, Sakyo-ku, Kyoto 606-8501, Japan

\setlength{\baselineskip}{0.77cm}
%\maketitle

\newpage
%\begin{article}

\begin{abstract}
\setlength{\baselineskip}{0.77cm} 
The locomotion of \textit{Caenorhabditis elegans} exhibits complex patterns. In particular, the worm combines mildly curved runs and sharp turns to steer its course. Both runs and sharp turns of various types are important components of taxis behavior. The statistics of sharp turns have been intensively studied. However, there have been few studies on runs, except for those on klinotaxis (also called weathervane mechanism), in which the worm gradually curves toward the direction with a high concentration of chemicals; this phenomenon was discovered recently. We analyzed the data of runs by excluding sharp turns. We show that the curving rate obeys long-tail distributions, which implies that large curving rates are relatively frequent. This result holds true for locomotion in environments both with and without a gradient of NaCl concentration; it is independent of klinotaxis. We propose a phenomenological computational model on the basis of a random walk with multiplicative noise. The assumption of multiplicative noise posits that the fluctuation of the force is proportional to the force exerted. The model reproduces the long-tail property present in the experimental data.
\end{abstract}


\newpage

\section*{INTRODUCTION}

The soil nematode \textit{Caenorhabditis elegans} (\textit{C.~elegans})
is unique in that
its 302 neurons and their network have been fully identified
\cite{White1986}.  In addition, the entire genomic sequence of
\textit{C.~elegans} has been determined
\cite{C_elegans_Sequencing_Consortium1998}.
On agar surfaces, the worm lies on one of 
its two sides and exhibits complex locomotion by crawling forward and backward.  
Complex locomotion is particularly evident when the worm responds to various types of
sensory inputs. 
Examples of such functional
responses include chemotaxis, thermotaxis, and isothermal
tracking. The identification of specific neurons involved
in these and other responses 
has been a topic of intensive research
\cite{Bargmann1991,Bargmann1993,Mori1995,Gray2005,Iino2009}.

The complex locomotion of \textit{C.~elegans},
which underlies functional responses of the worm,
 accompanies
undulations of the entire body with a period of 2--3 s.
In general, trajectories of the worm are not straight but 
mildly curved. The worm often turns
sharply. Sharp 
turning events are called under different names,
such as reversals, sharp turns, and omega turns, depending on
the specific manner of turning.
A bout of these sharp turning events is generally
called a pirouette, and pirouette is known to
improve the efficiency of the worm's locomotion toward
chemoattractants (chemotaxis) \cite{Pierce-Shimomura1999}.
Computational modeling has been a useful tool for understanding
the locomotion.
Most existing computational models 
assume that 
sharp turning events, or pirouettes, are
the main drive of the functional locomotion behavior
of the worm
\cite{Pierce-Shimomura1999,Matsuoka2008,Ramot08jns,Nakazato09}.

Recently, another mechanism of chemotaxis was discovered \cite{Iino2009}. 
It was
called weathervane mechanism in \cite{Iino2009}, but more generally called
klinotaxis \cite{Frankel_book}.
In klinotaxis in \textit{C.~elegans}, an animal performing a run (segment of
trajectory between adjacent sharp turns; see MATERIALS AND METHODS for
definition) senses the spatial gradient of chemical compounds in the
direction lateral to its forward movements and gradually curves toward a
higher concentration of the chemicals. This is distinct from the
pirouette mechanism in which pirouettes occur frequently when the
concentration of the attractant progressively decreases \cite{Pierce-Shimomura1999}, which is a
form of klinokinesis \cite{Frankel_book}. Klinotaxis suggests that
modeling the trajectory of the worm by the combination of pirouettes and
straight runs, which do not allow for klinotaxis, may be an
oversimplification. There have been few computational studies of runs,
except for some on klinotaxis \cite{Iino2009} and isothermal tracking \cite{Luo06}.

In this paper, we analyze run data of the locomotion of \textit{C.~elegans} 
in either the absence or presence of a chemoattractant in detail. 
Our main finding is that the curving angle per unit time exhibits a long-tail distribution
 (the precise definition of the curving angle is given in MATERIALS AND METHODS). 
This finding is robust against the influence of the attractant,
laser ablation of neurons, and bias in the curving inherent in each worm.  
Then, we propose a minimal computational model 
using a correlated random walk, which has been used for modeling 
the migration of insects and
mammals \cite{Siniff69,Hall77,Kareiva83,Bartumeus05ecol,Codling08}.  
In contrast to these previous cases, the increment of
the curving angle cannot be modeled as a Gaussian distribution,
which would arise as a summation of many independent
and identically distributed noisy microscopic movements
 \cite{Kitching71,Bovet88jtb}.
A random walk with 
multiplicative random noise \cite{Takayasu97} explains the long-tail
behavior in the locomotion of \textit{C.~elegans}.

\section*{MATERIALS AND METHODS}

\subsection*{Experimental details}

We analyzed a subset of 
the experimental data used in ref.~\cite{Iino2009}.
We briefly explain the experimental setup and the data 
(see ref.~\cite{Iino2009} for details).

We used data from two sets of experiments.  
In one environment, each worm was placed on an agar plate without NaCl.
This data set is the main focus of the present study.
%We recorded the locomotion of 46 worms in this environment.
Data obtained from 46 worms in this environment were analyzed.
In the other environment, we placed the worms on a grid format plate in which NaCl was
spotted in a grid pattern and allowed to diffuse on chemotaxis agar as follows.
%We recorded the locomotion of 53 worms in this environment.
Data obtained from 53 worms in this environment were analyzed.

Worms (\textit{C.~elegans}; Bristol strain N2) were cultured
at 20\degC using a standard technique.
The agar plates contained 10 ml of 1 mM CaCl$_2$,
1 mM MgSO$_4$, 5 mM potassium phosphate, pH 6.0, and 2\% agar
in a 9-cm-diameter plastic dish.
For assays with NaCl (chemotaxis assays), the agar plates were prepared as follows:
1 $\mu$M each of 200 mM NaCl in a low-salt buffer was spotted
onto 12 points in a grid, which were spaced 2 cm apart;
the positions of the centers of the spots were $(x,y)$ = (10 mm, 10 mm), (10 mm, $-$10 mm), 
($-$10 mm, 10 mm), and so on (see \cite{Iino2009}).
The assay plates were left for 1 h
after spotting the NaCl-containing buffer.

In the tracking system, images were captured and analyzed to locate
the centroid of the worm.  
Images were captured at intervals 0.4--0.6 s.  
Each worm was tracked for 20 min after being placed on the
agar plate.  Tracking data were discarded when a worm was considered
to be immobile during an entire recording session.

\subsection*{Data analysis}

%\begin{figure}
%\begin{center}
%\includegraphics[width=70mm]{fig_real_sample_trajectory}
%\includegraphics[width=55mm]{fig_various_def}
%\includegraphics[width=45mm]{fig_def_sharp_turn}
%\caption{
%({\it A}) Sample trajectory of the worm.
%(B) Definition of the curving rate $\phi_i$.
%(C) Definition of the sharp turn segment (shaded area).
%}
%\label{fig_def_various_quantities}
%\end{center}
%\end{figure}

\begin{figure}[h]
\begin{center}
\includegraphics[width=70mm]{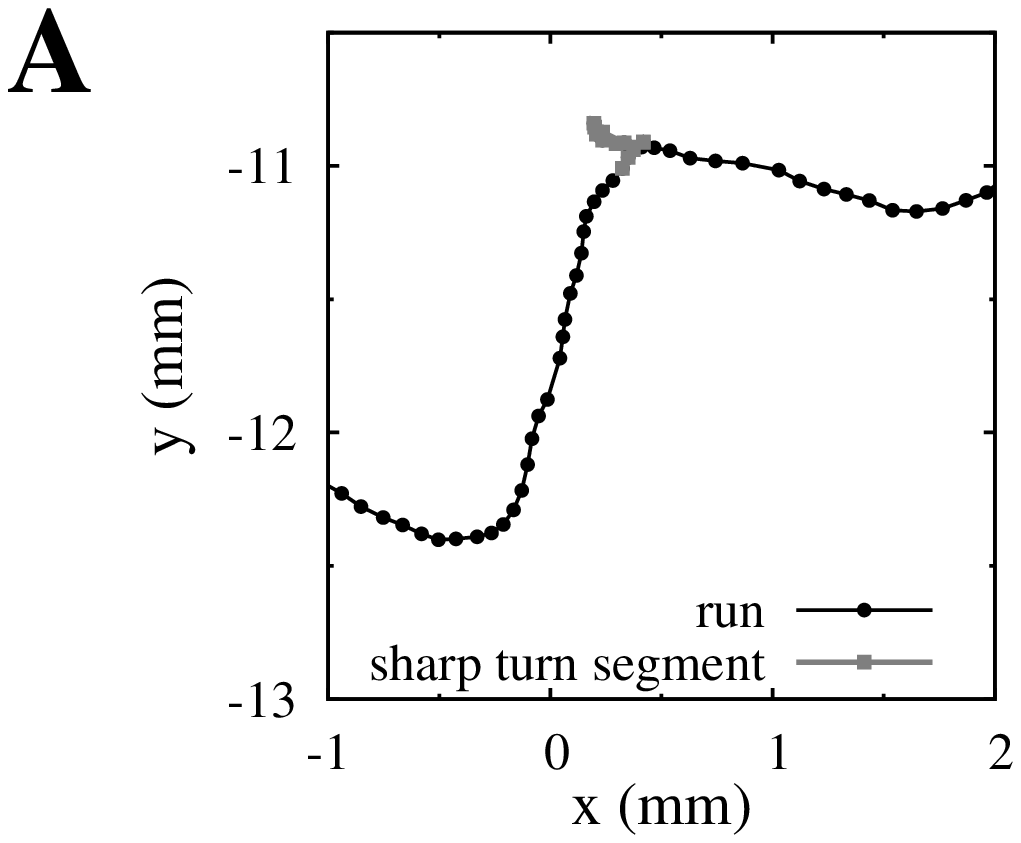}
\hspace{5mm}
\includegraphics[width=55mm]{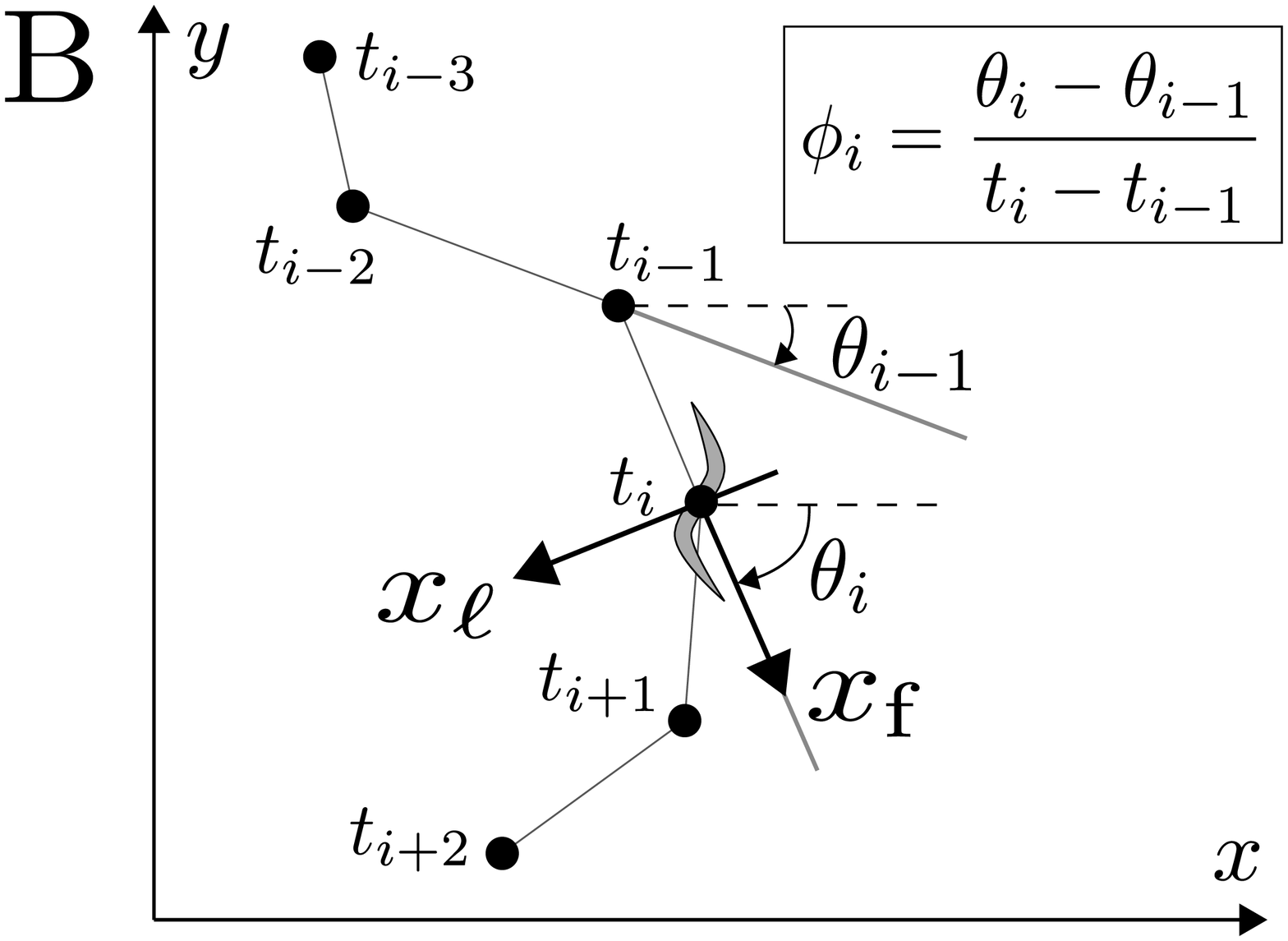}\\

\vspace{15mm}
\includegraphics[width=45mm]{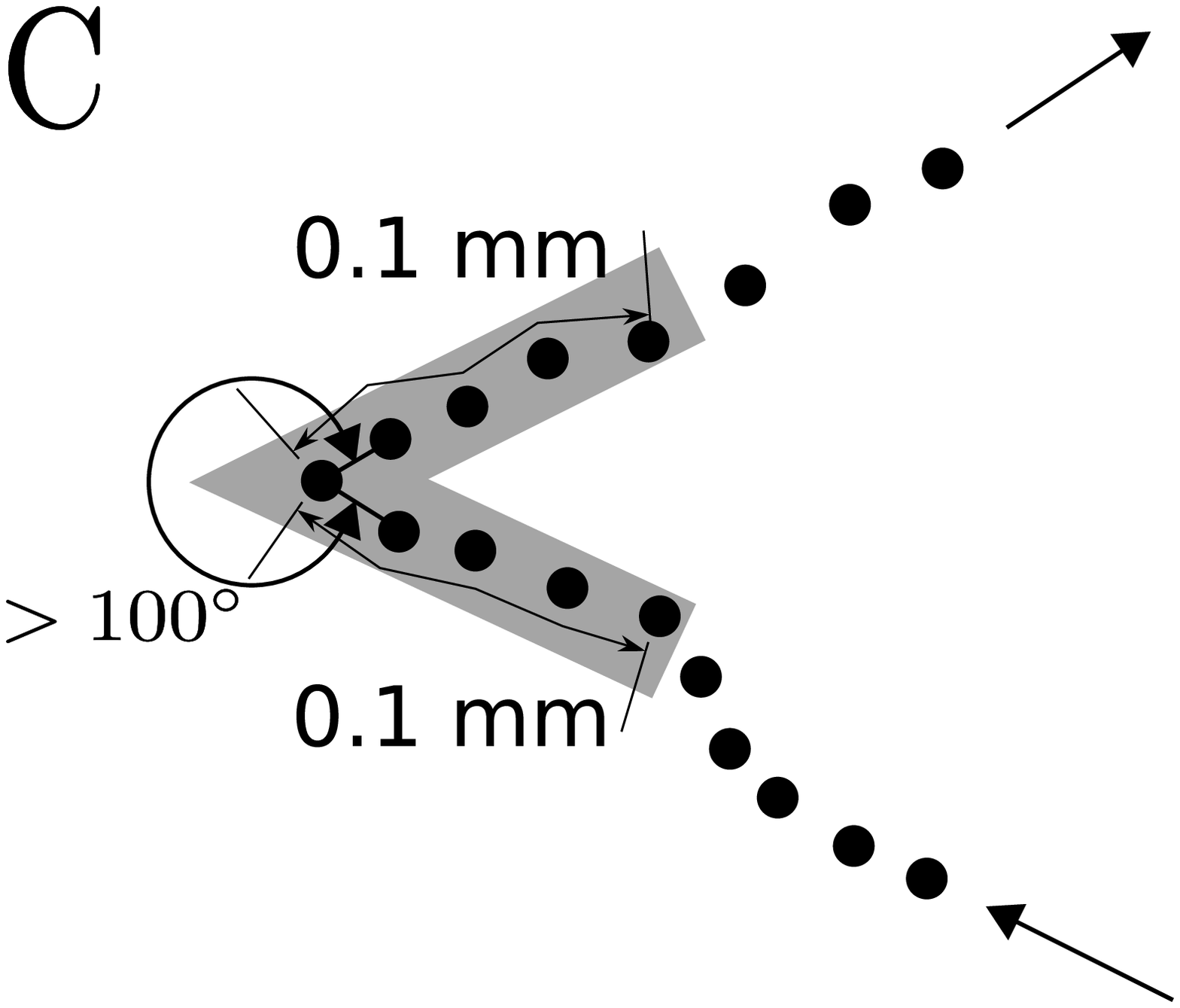}
\caption{
%(A) Sample trajectory of the worm.
%(B) Definition of the curving rate $\phi_i$.
%(C) Definition of the sharp turn segment (shaded area).
}
\label{fig_def_various_quantities}
\end{center}
\end{figure}

A sample trajectory of a worm in the two-dimensional plane is shown in
Fig.~\ref{fig_def_various_quantities}A.
The symbols represent the location at each image capture.
We discard sharp turns (gray squares)
defined in the following and analyze
the rest of the data (dark circles).
To this end, we define the displacement vector as
the difference between the coordinate observed at a time point and
that observed at the previous time point
(Fig.~\ref{fig_def_various_quantities}B).  
The angle from the $x$-axis
to the displacement vector at time $t_i$ 
measured clockwise is 
denoted as $\theta_i$.
%\edit{For example, $\theta_{i-1}$ and $\theta_i$
%in Fig.~\ref{fig_def_various_quantities}B are defined to be negative.}
Following \cite{Pierce-Shimomura1999}, 
we define the curving rate as
\begin{equation}
\phi_i = \frac{\theta_i - \theta_{i-1}}{t_i-t_{i-1}}.
\end{equation}
This definition is slightly different
from that in \cite{Iino2009}, in which the curving rate is defined by
the derivative of $\theta$ with respect to the distance traveled.  The
determination of a curving rate involves three successive data points.
The speed in addition to the curving rate is required to specify 
the locomotion of the worm. Although the speed of the
worm varies with time, we neglect these variations, as in the case of
previous modeling studies \cite{Pierce-Shimomura1999,Iino2009,Dunn2004}

We remove sharp turns as follows
(Fig.~\ref{fig_def_various_quantities}C).
A sharp turn is defined as an event
in which $|\theta_{i+1} - \theta_i| > 100^\circ$ 
is satisfied.
A sharp turn segment (shaded area in
Fig.~\ref{fig_def_various_quantities}C)
is defined as the segment of the 
trajectory within 0.1 mm
from a sharp turn event. The length of the trajectory is measured by
the accumulated length of the piecewise linear path connecting
the two-dimensional coordinates.
We remove all the sharp turn segments and analyze the rest of the data,
which we call the run data. 
A run is a section of the trajectory between two successive
sharp turn segments. A single worm generates a trajectory that generally
includes multiple runs.

Our data were recorded at irregular intervals.
To avoid the technical complication of calculating the autocorrelation of
the curving rate in this situation, we proceed as follows.
We first generate a continuous
trajectory on the two-dimensional plate 
from a sequence of the coordinates of the worm
by using the third spline interpolation (numerically
implemented using GNU Scientific Library). 
Then, we resample discrete time series of the coordinates
from the interpolated continuous trajectory
every 0.3 s. Next, we detect sharp turns of the resampled trajectory
and exclude the sharp turn segments.
$\phi_{s,i,j}$, $\ell_{s,i}$, and $M_s$ denote the $j$th curving rate
in the $i$th run of the $s$th worm ($1\le s\le 46$),
length of the $i$th run of the
$s$th worm, and number of runs of the $s$th worm, respectively.
We combine all the
runs from all the worms to calculate a single autocorrelation function
because reliably calculating the
autocorrelation necessitates many data points.

The number of data points used for calculating the
autocorrelation function with lag $\tau$ is given by
\begin{align}
L_\tau = \sum_{s=1}^{46} \sum_{i=1}^{M_s} \max(\ell_{s,i}-\tau, 0).
\end{align}
The average curving rate for the data points that are included
in the calculation of the two-point 
correlation function with lag $\tau$ is equal to 
\begin{align}
\left<\phi\right>_\tau \equiv
\frac{1}{L_\tau} \sum_{s=1}^{46} \sum_{i=1}^{M_s} 
\sum_{j=1}^{\max(\ell_{s,i}-\tau,0)} \phi_{s,i,j}.
\end{align}
The autocorrelation function with lag $\tau$ (i.e., $0.3\times \tau$ s),
denoted by $R_\tau$, is defined by
\begin{align}
R_\tau = \frac{\displaystyle{\frac{1}{L_\tau} \sum_{s=1}^{46} \sum_{i=1}^{M_s}
\sum_{j=1}^{\max(\ell_{s,i}-\tau,0)}
\left(\phi_{s,i,j} - \left<\phi\right>_\tau \right)
\left(\phi_{s,i,j+\tau} - \left<\phi\right>_\tau \right)}}
{\displaystyle{\sqrt{
\frac{1}{L_\tau} \sum_{s=1}^{46} \sum_{i=1}^{M_s} \sum_{j=1}^{\max(\ell_{s,i}-\tau,0)} 
 \left(\phi_{s,i,j} - \left<\phi\right>_\tau \right)^2
}
{\sqrt{
\frac{1}{L_\tau} \sum_{s=1}^{46} \sum_{i=1}^{M_s} \sum_{j=1}^{\max(\ell_{s,i}-\tau,0)} 
 \left(\phi_{s,i,j+\tau} - \left<\phi\right>_\tau\right)^2
}}}}.
\end{align}
Note that $R_0=1$.
We verified that the interpolation does not significantly
affect the trajectory or the distributions of $\phi$. 
We employed the spline interpolation only for calculating the
autocorrelation function.

When NaCl is present, the concentration of
NaCl created by a single spot of NaCl was evaluated by the solution of
the diffusion equation: $N_0 \exp(-r^2/(4 Dt)) / (4 \pi d D t)$,
where $N_0 = 2.0\times 10^{-4}$ (mM) is the amount of NaCl spotted, $D 
=
1.5\times 10^{-3}$ (mm$^2$/s) is the diffusion constant of NaCl, $d =
1.8$ (mm) is the thickness of the plate, $t$ is the time (in seconds) after
spotting, and $r$ is the distance (in millimeters) 
from the spot. 
The summation of the estimated NaCl
concentration based on 12 NaCl spots 
defines the total concentration of NaCl, which we denote by $C$.

To examine klinotaxis, we compute the gradient of the
concentration of NaCl along two directions. One is the forward
direction, which is parallel to the displacement vector. The corresponding
gradient is denoted by $\mathrm{d}C/\mathrm{d} x_{\mathrm f}$. 
The other is the lateral direction, which is
perpendicular to the forward direction.  
The corresponding gradient is
denoted by $\mathrm{d}C/\mathrm{d} x_{\mathrm \ell}$.
We define $\mathrm{d}C/\mathrm{d} x_{\mathrm \ell}$ to be
positive if $C$ on the right side
of the worm, when viewed from above, is larger than that on the left side.
%\edit{Here, the lateral direction is positive for the rightward direction along 
%the displacement vector,
%as shown in Fig.~\ref{fig_def_various_quantities}B.}

\section*{RESULTS}

\subsection*{Long-tail behavior of curving rate}

%\subsubsection*{\edit{In absence of NaCl gradient}}

We analyze runs from two sets of experiments from different environments:
ones 
with and without a gradient of the NaCl concentration (NaCl gradient
for short). We mainly focus on the data recorded
in the absence of an NaCl gradient.

%\begin{figure}
%\begin{center}
%\includegraphics[width=70mm]{fig_real_sample_path.eps}
%\includegraphics[width=70mm]{fig_real_auto_correlation.eps}
%\includegraphics[width=70mm]{fig_real_distribution.eps}
%\includegraphics[width=70mm]{fig_real_cumulative_distribution.eps}
%\includegraphics[width=70mm]{fig_no_grad_dist_sharp_turn.eps}
%\caption{Results obtained from experiments in the absence of
%an NaCl gradient.
%({\it A}) Sample trajectory of $\phi$.
%(B) Autocorrelation function of $\phi$.
%(C) Distributions of $\phi$
%based on the runs of a single worm and those of all the worms.
%We use a bin width of 4\degs.
%(D) Cumulative distributions of $|\phi|$ on the log-log scale.
%The results for one worm and the combined one for all the worms are shown
%separately for positive and negative values of $\phi$.
%The dot-dashed line in (D) indicates
%the cumulative distribution of the Gaussian distribution with mean
%zero and standard deviation 15.8. 
%(E) Distribution of $|\phi|$ for the data
%before and after sharp turns are excluded. The data for all the worms are
%combined to generate each distribution.}
%\label{fig phi no-NaCl}
%\end{center}
%\end{figure}

\begin{figure}
\begin{center}
\includegraphics[width=70mm]{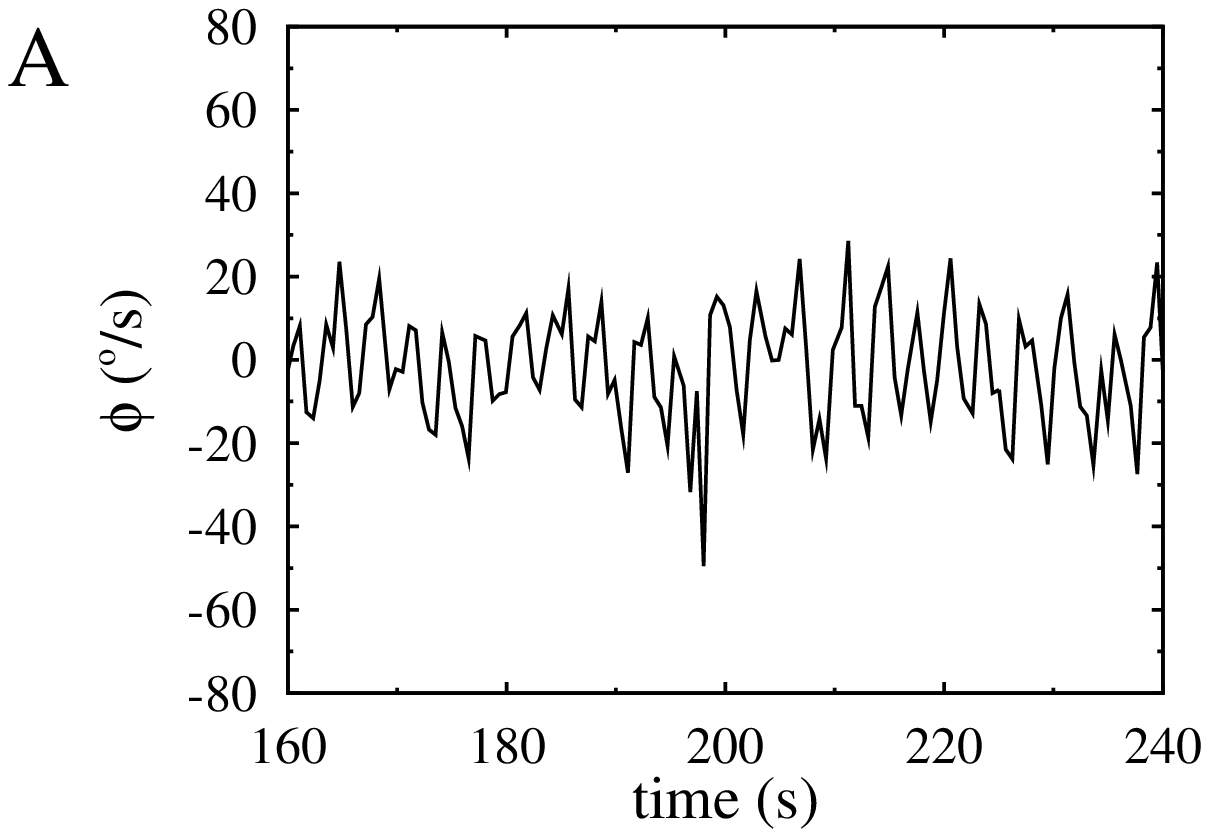}
\includegraphics[width=70mm]{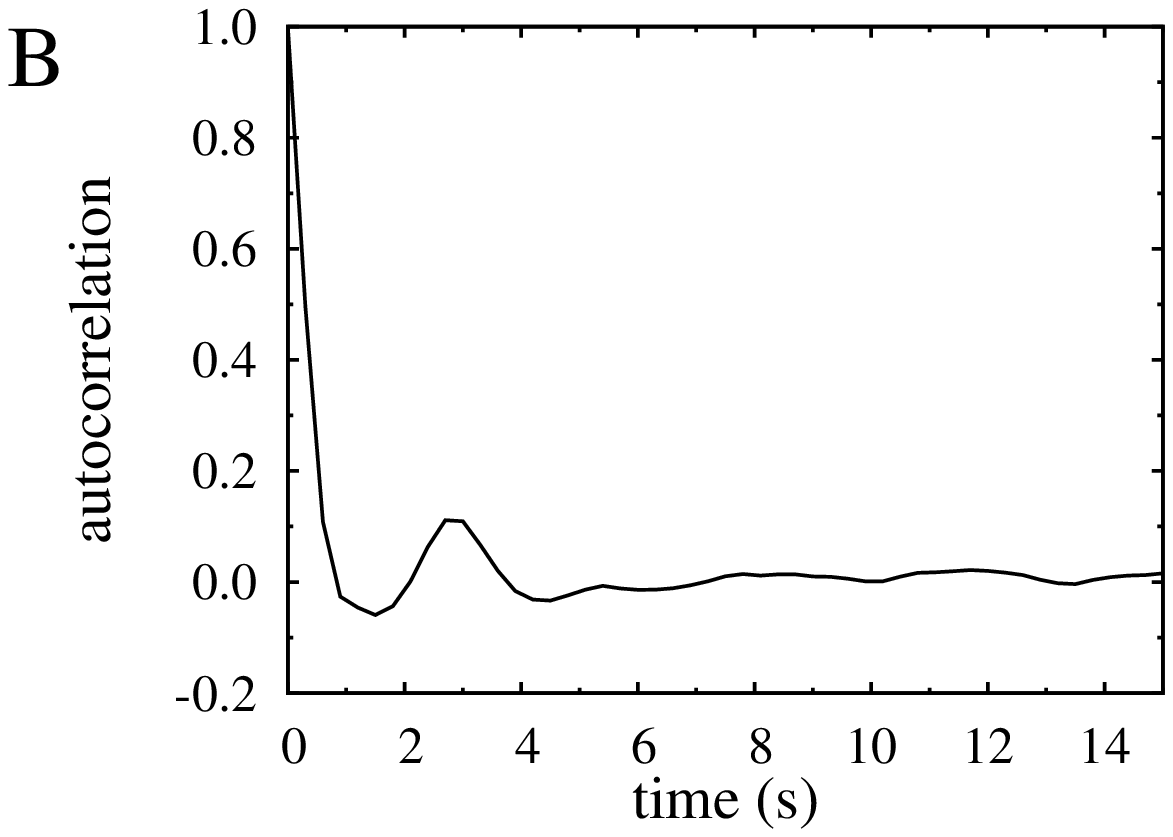}
\includegraphics[width=70mm]{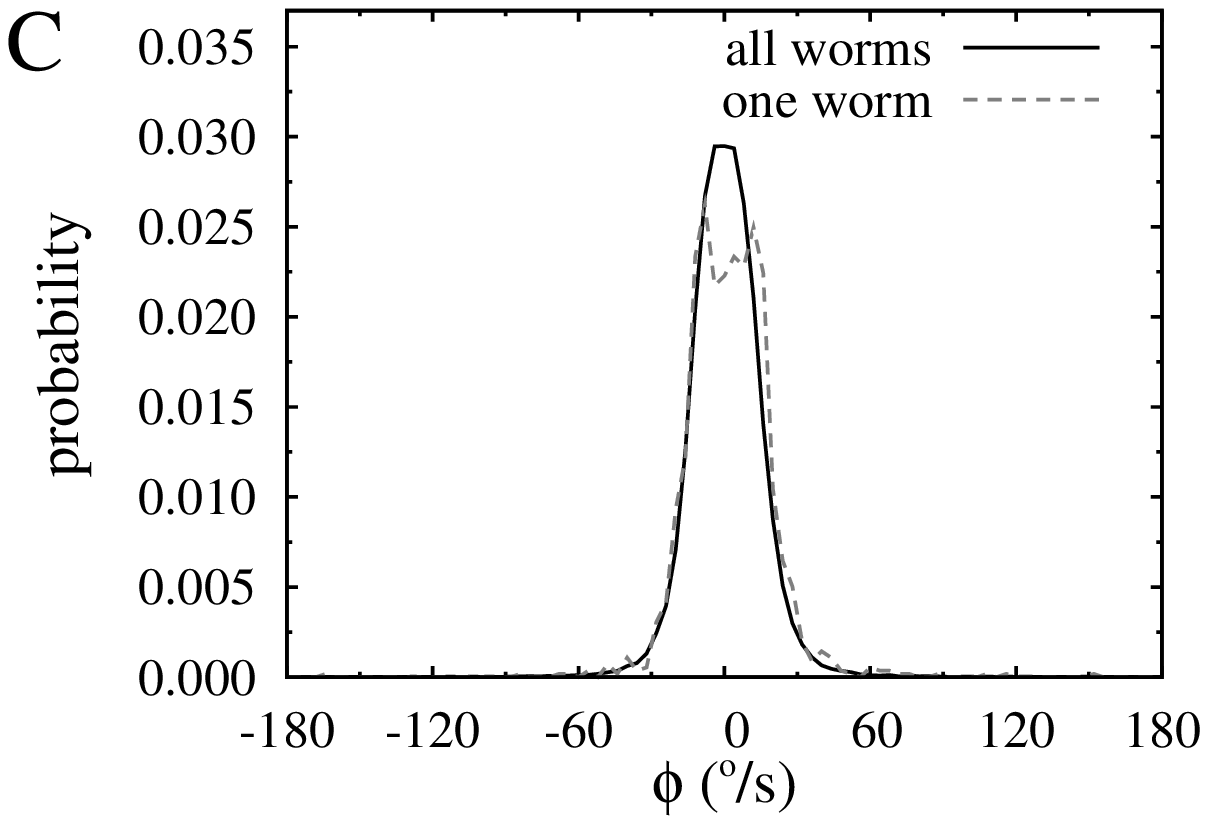}
\includegraphics[width=70mm]{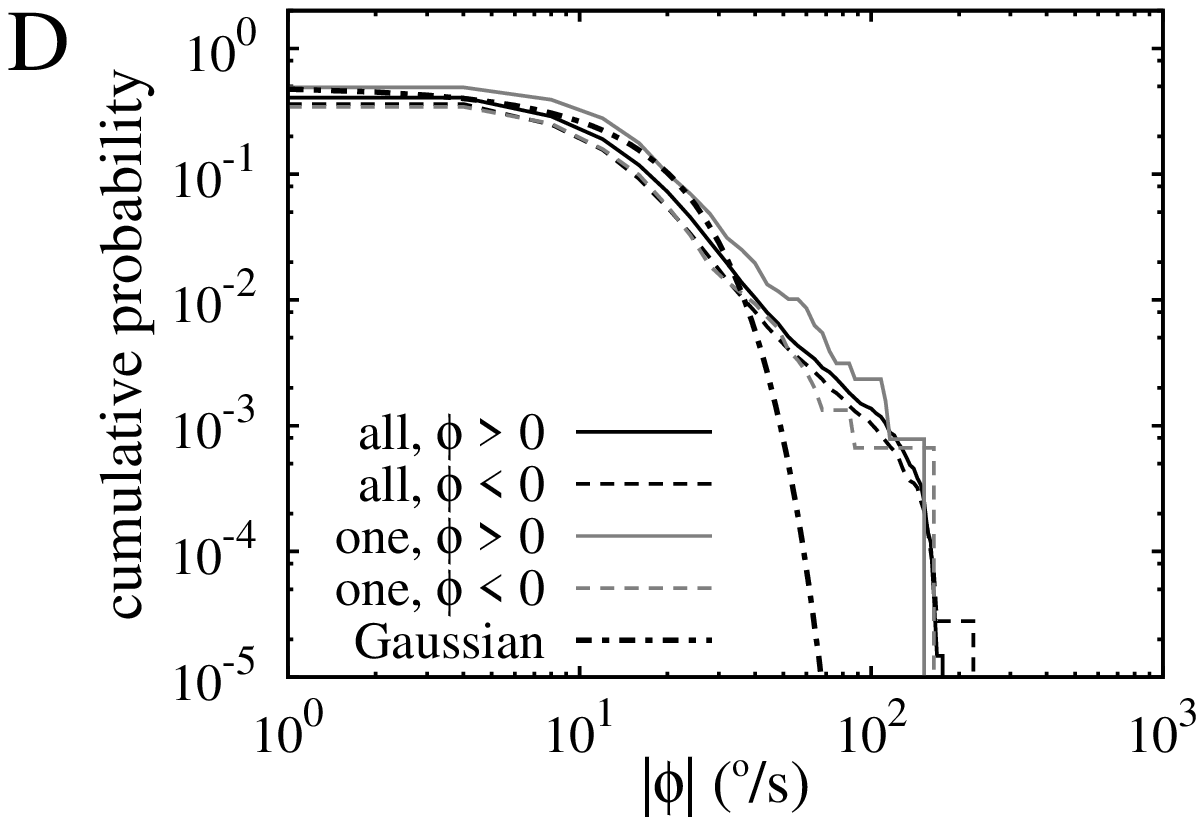}
\includegraphics[width=70mm]{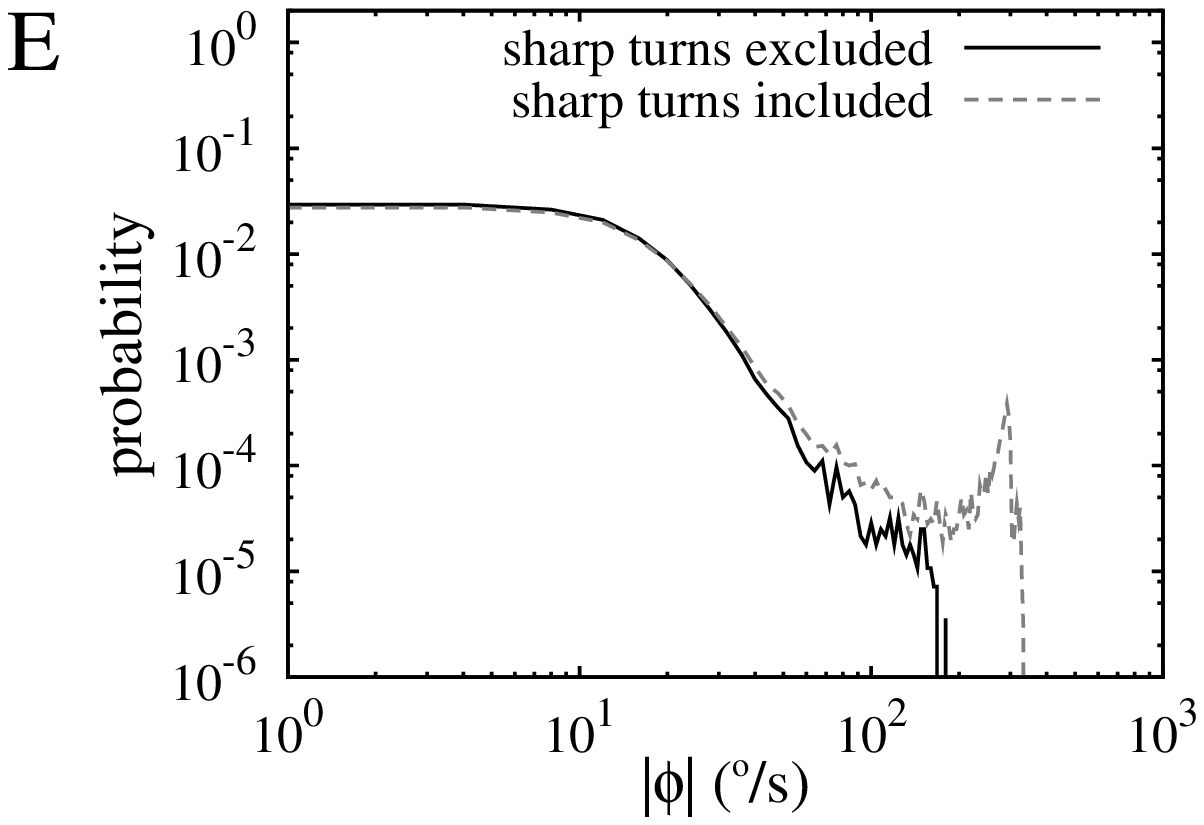}
\caption{
%Results obtained from experiments in the absence of
%an NaCl gradient.
%(A) Sample trajectory of $\phi$.
%(B) Autocorrelation function of $\phi$.
%(C) Distributions of $\phi$
%based on the runs of a single worm and those of all the worms.
%We use a bin width of 4\degs.
%(D) Cumulative distributions of $|\phi|$ on the log-log scale.
%The results for one worm and the combined one for all the worms are shown
%separately for positive and negative values of $\phi$.
%The dot-dashed line in (D) indicates
%the cumulative distribution of the Gaussian distribution with mean
%zero and standard deviation 15.8. 
%(E) Distribution of $|\phi|$ for the data
%before and after sharp turns are excluded. The data for all the worms are
%combined to generate each distribution.
}
\label{fig phi no-NaCl}
\end{center}
\end{figure}

A time series of the curving rate $\phi$ of a worm during a run is
shown in Fig.~\ref{fig phi no-NaCl}A.  $\phi$ has a periodic
component and accompanies large fluctuations.  
Periodic behavior with a period of
approximately 3 s is derived from the body undulation.
The autocorrelation
function of the curving rate shown in Fig.~\ref{fig phi no-NaCl}B
indicates oscillatory damping.

The distribution of $\phi$ for a typical worm is indicated by the
dashed line in Fig.~\ref{fig phi no-NaCl}C.  The distribution
has two peaks owing to the oscillatory undulation.  The significance
of the two peaks varies between worms and runs.  When we combine the
data of all the worms, the distribution of $\phi$ is unimodal (solid
line in Fig.~\ref{fig phi no-NaCl}C), which reflects the
heterogeneity of the worms. To inspect the tails of the distribution,
we divide the samples into a group with $\phi>0$ and that with
$\phi<0$.  For a single worm, the cumulative distributions of $|\phi|$,
defined in the present study as the fraction of samples among 
all the samples such that the absolute value is larger than $|\phi|$, are separately shown 
for the two groups in
Fig.~\ref{fig phi no-NaCl}D.  
Remarkably, both two cumulative
distributions of $\phi$ have long tails.  
The long tails are also
evident when we combine the data of all the worms (Fig.~\ref{fig phi
no-NaCl}D).  The mean of $\phi$ for the combined data is almost
equal to zero, and the standard deviation of $\phi$ is equal to
$15.8$.  The cumulative distribution of the Gaussian distribution with
mean zero and standard deviation 15.8 is indicated by the dot-dashed
line in Fig.~\ref{fig phi no-NaCl}D.  The distribution of
$\phi$ is not accurately fitted by the Gaussian distribution because
of the long tail present in the experimental data.  A manual power-law
fit to the tail of the cumulative distribution of $|\phi|$ yields the
power law $\propto |\phi|^{-2.5}$ (i.e., $\propto |\phi|^{-3.5}$ for
the original distribution).  The exponent and the significance of the
power law depend on the worm and on the sign of $\phi$.  Nevertheless,
for all the data examined, the tail is much longer than that of the
Gaussian distribution of the same value of the standard deviation.

To show that the observed long-tail distribution of $\phi$ is not
a byproduct, we perform a few tests.
First, the long tail is not caused by sharp
turns. If we include the sharp turns in the analysis, we find a peak
at a large value of $\phi$ (Fig.~\ref{fig phi no-NaCl}E) that
is distinct from the long tail observed in the run data. 
For visibility, the original distributions of $|\phi|$, not the
cumulative ones as in other similar figures, are plotted
in Fig.~\ref{fig phi no-NaCl}E.

Second, the long tail is not caused by
the immobility of the worm. 
The mean of $|\phi|$ is plotted 
against the instantaneous speed of the worm in
Fig.~\ref{fig: real
cumulative phi for verification}A.
%in Supporting Material. 
$|\phi|$ tends to be large
when the worm runs slowly.
Excluding the values of $\phi$
for which the instantaneous
speed of the worm is small, i.e., less than 0.05 mm/s,
preserves the long-tail distribution (Fig.~\ref{fig: real
cumulative phi for verification}B). 

\begin{figure}[h]
\begin{center}
\includegraphics[width=70mm]{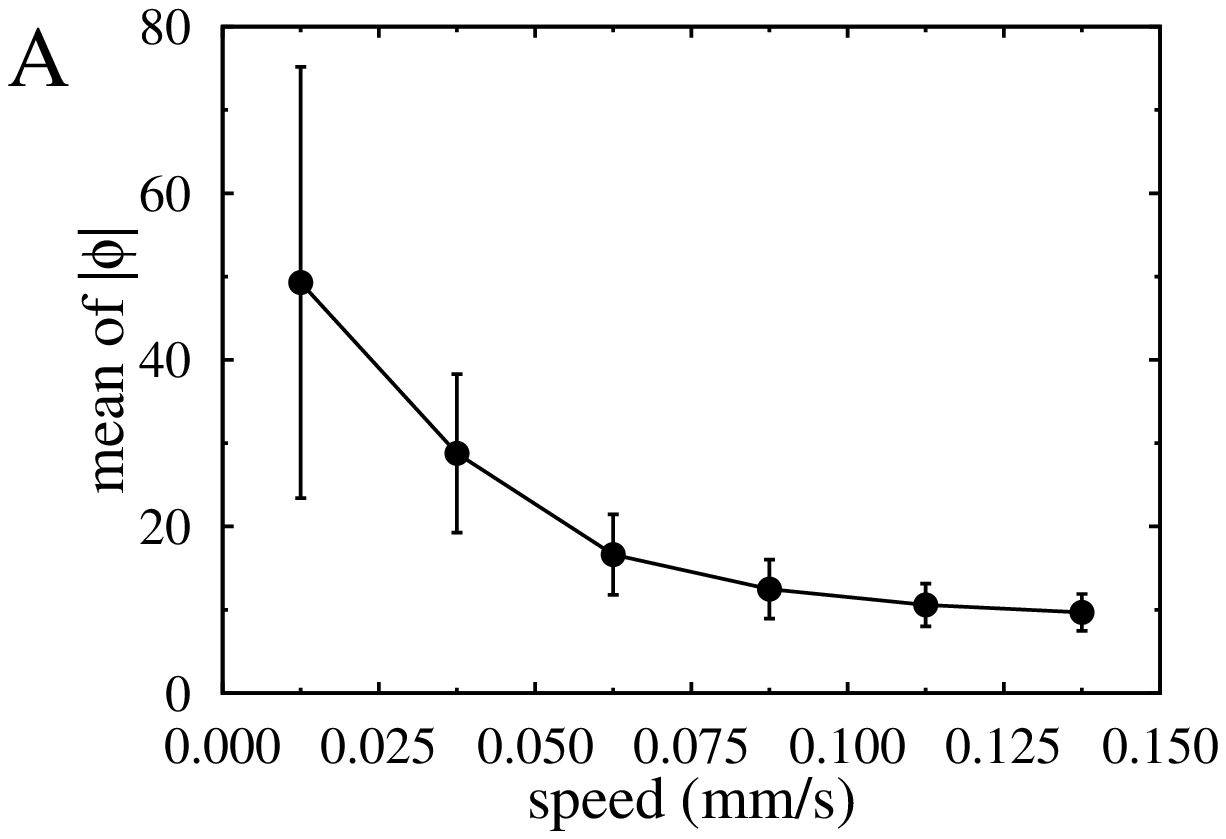}
\includegraphics[width=70mm]{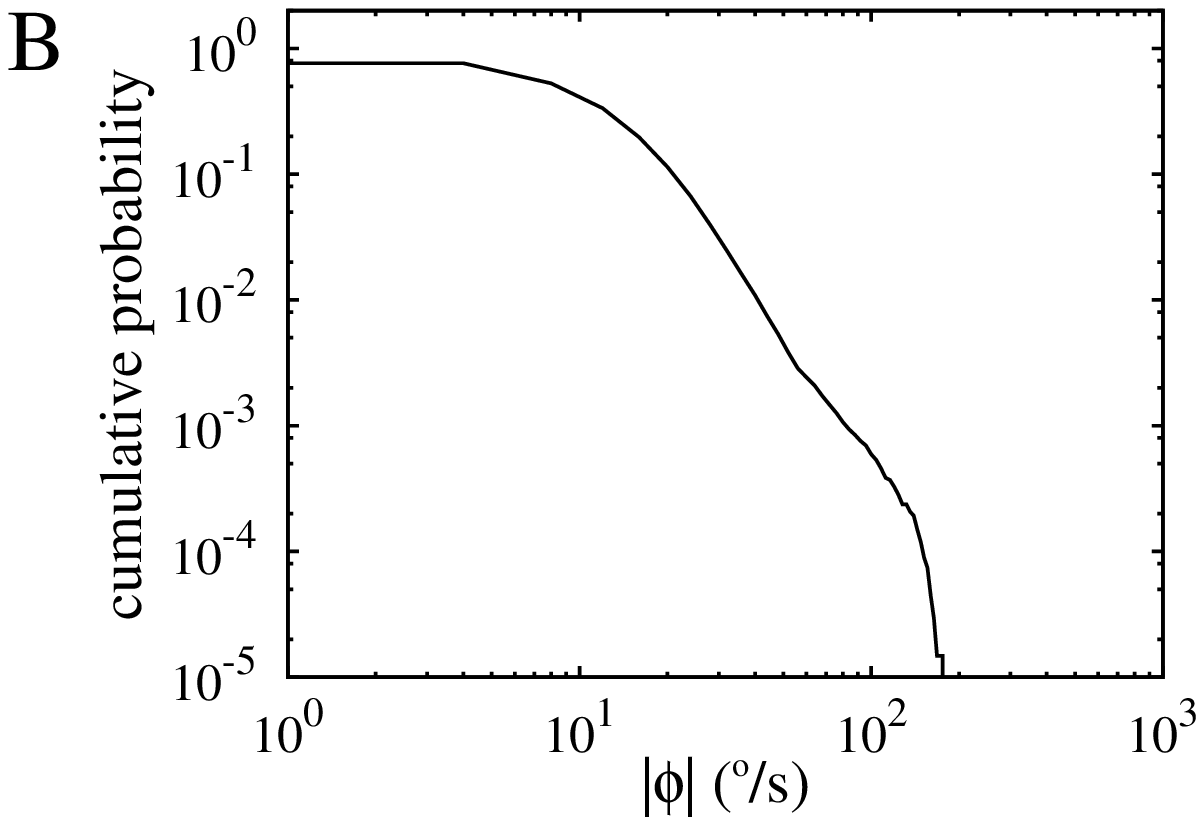}
\includegraphics[width=70mm]{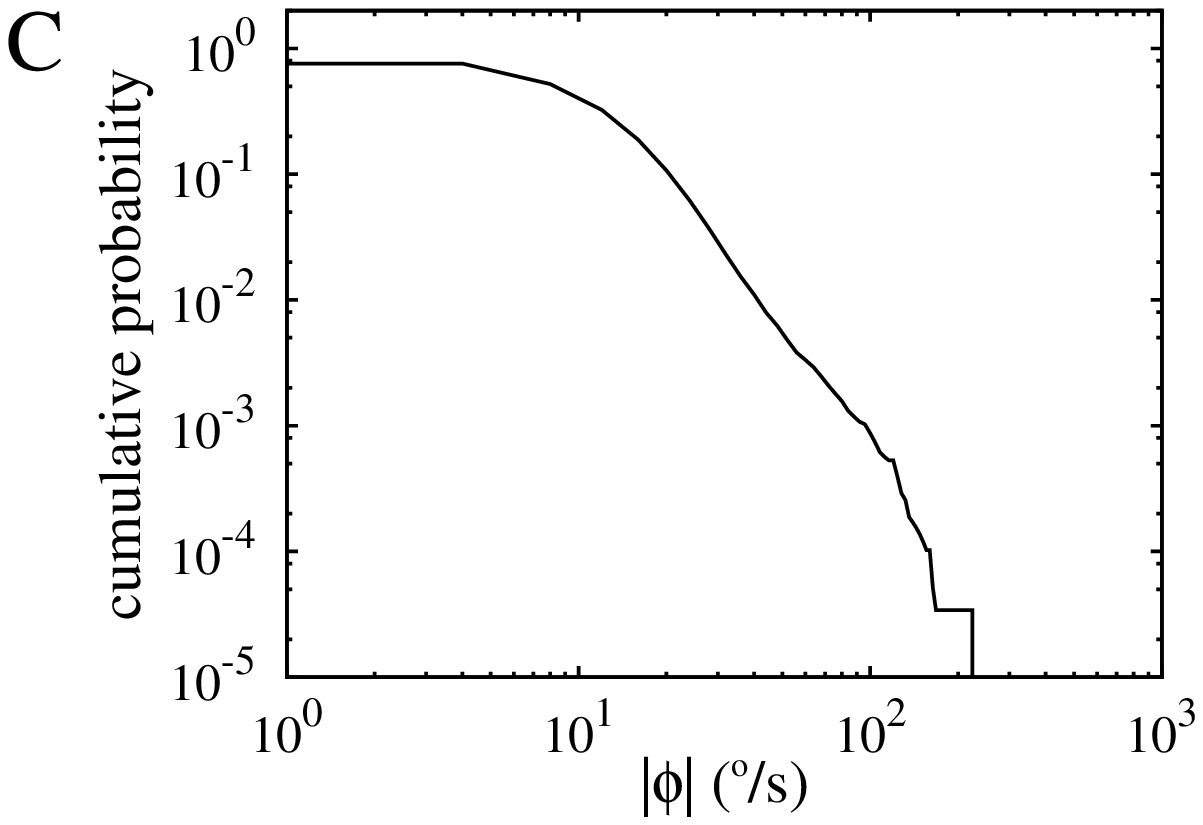}
\caption{
% (A) Relationship between the 
% mean of $|\phi|$ and the
% instantaneous speed of the worms. The mean values are
% calculated from all the runs of a single worm
% by categorizing the data into bins of width 0.025 mm/s
% and taking the average in each bin.
% Then, the mean values from different worms for the same bin
% are used for calculating the grand mean and
% standard deviation. The grand mean and
% error bar defined by the one standard deviation range 
% are plotted in (A).
% (B) Cumulative distribution of $|\phi|$ when
% data points for which the speed of the worms is less than 0.05
% mm/s are excluded.
% (C) Cumulative distributions of $|\phi|$ when the
% data points within 6 s before or after each sharp turn are excluded.
% For each figure, the data obtained
% from all the worms are combined to generate the cumulative
% distribution.
}
\label{fig: real cumulative phi for verification}
\end{center}
\end{figure}

Third, the occurrence of
a large value of $|\phi|$ is not restricted to periods near
sharp turns. We recover the long-tail distribution even if
we exclude a significant length of run (i.e., 6 s) before and after sharp
turns (Fig.~\ref{fig: real cumulative phi for verification}C).

%\subsubsection*{\edit{In presence of NaCl gradient}}

\begin{figure}
\begin{center}
\includegraphics[width=70mm]{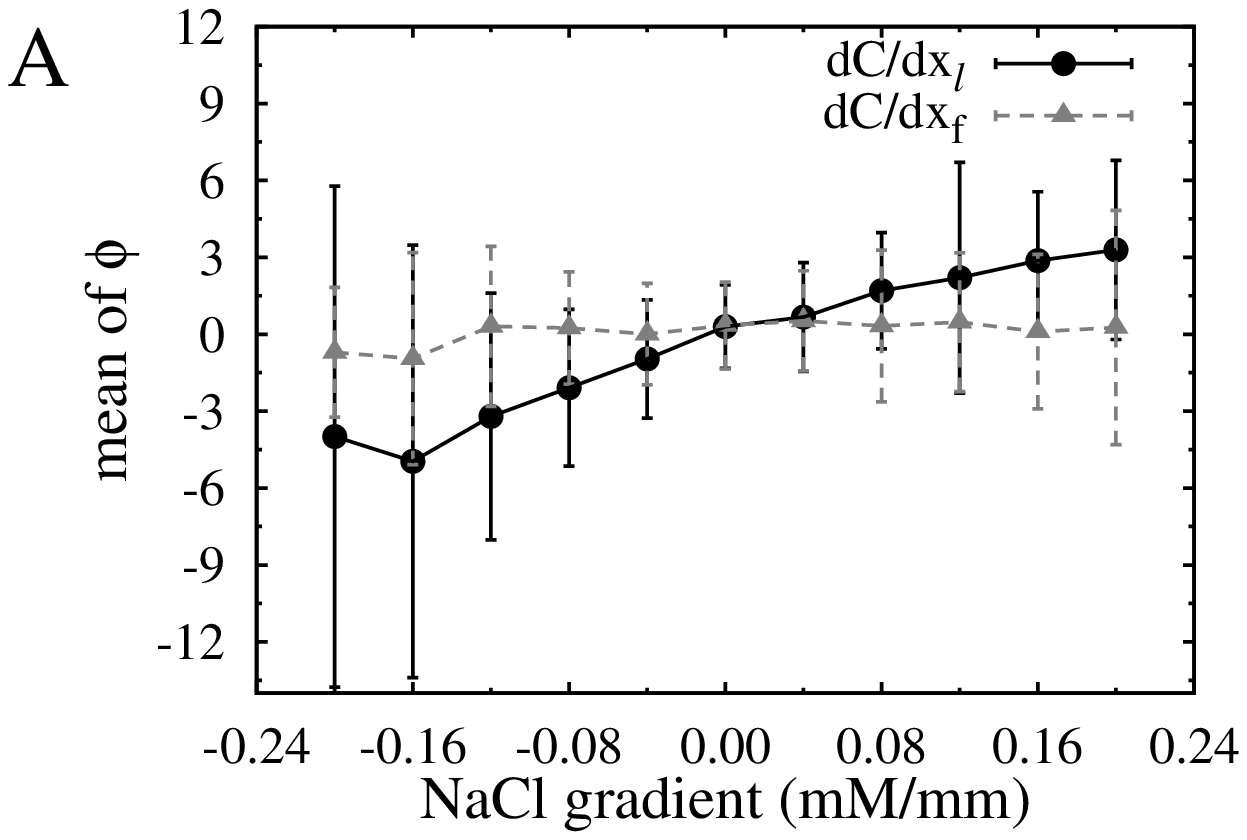}
\includegraphics[width=70mm]{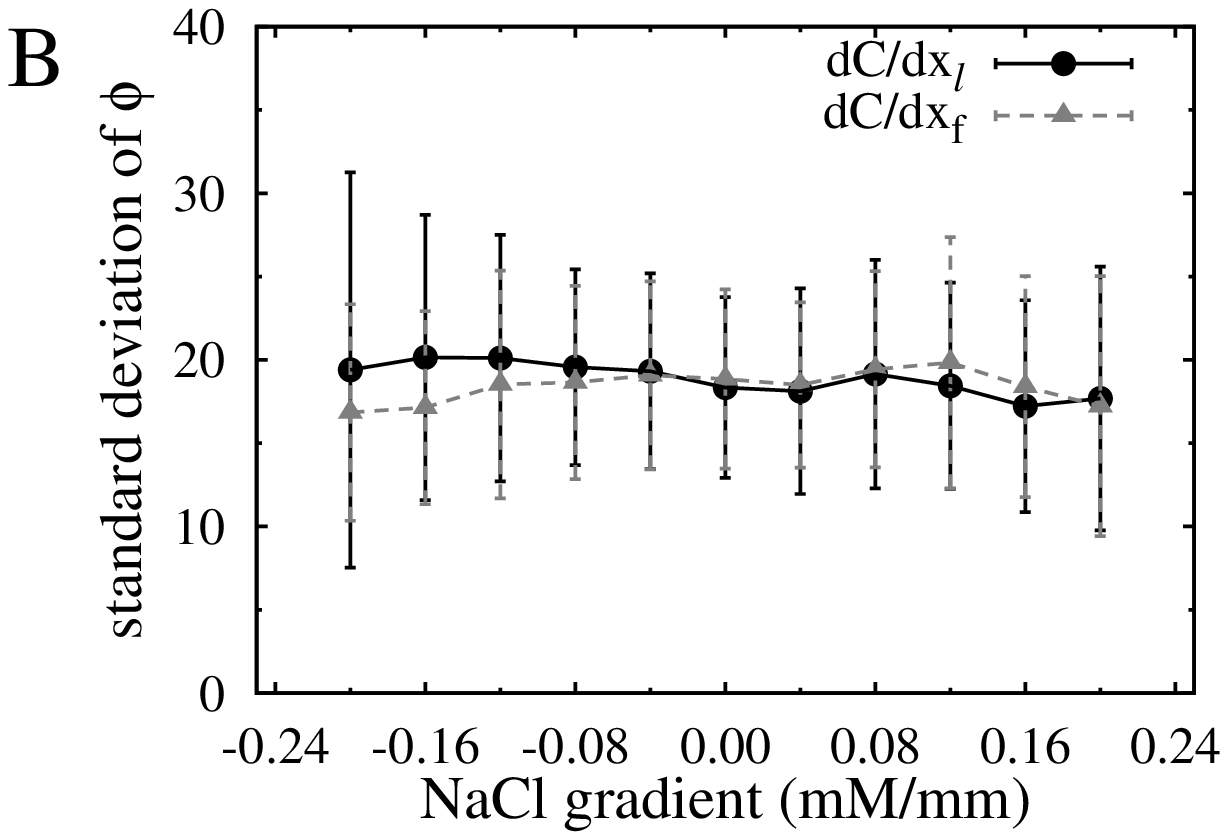}
\includegraphics[width=70mm]{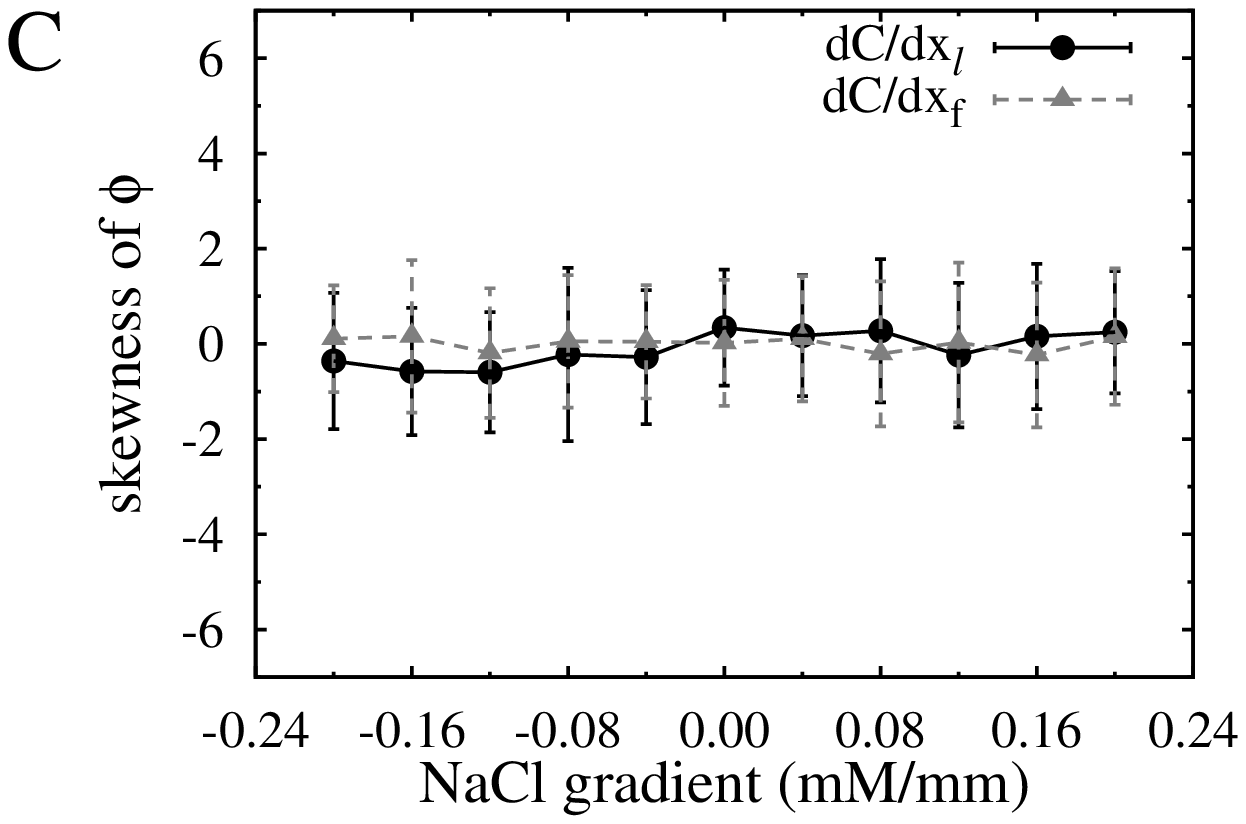}
\includegraphics[width=70mm]{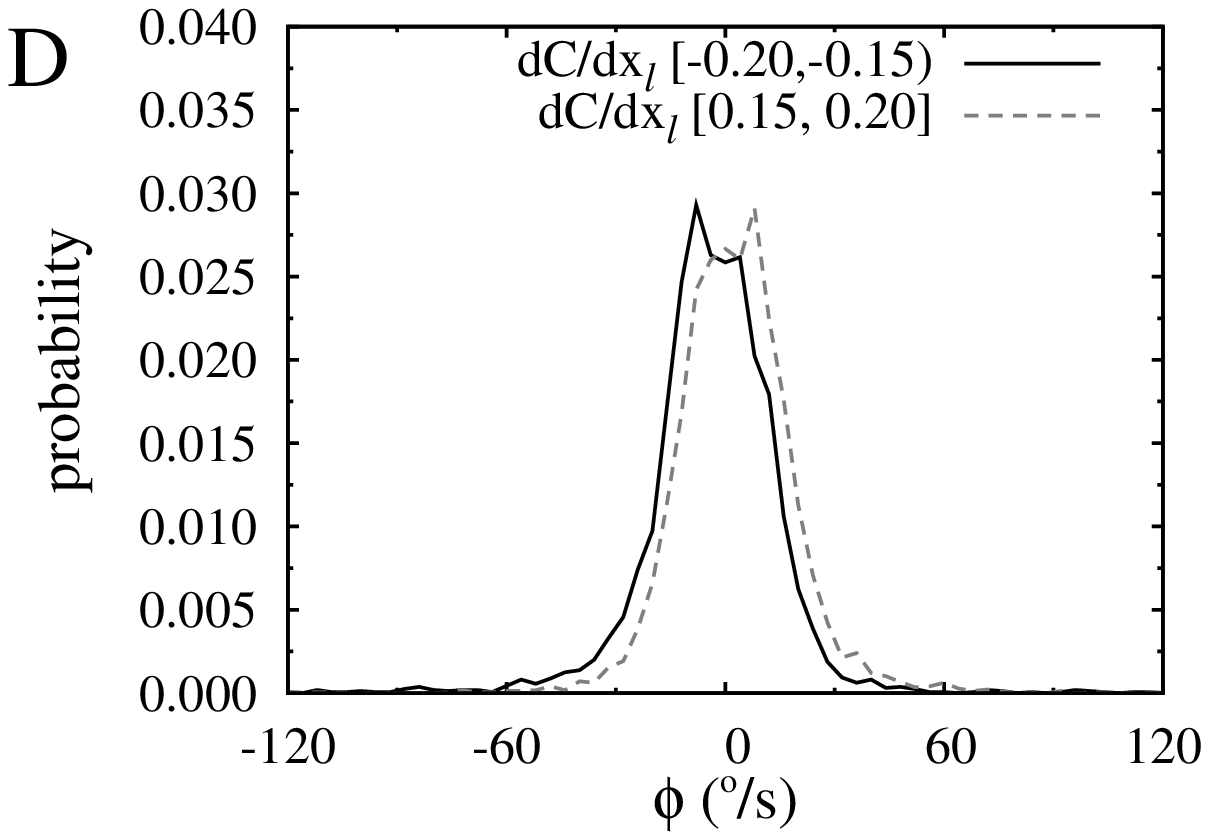}
\caption{
%Weathervane mechanism.
%(A) Mean, (B) standard deviation, and (C) skewness 
%of $\phi$ for different values of lateral and forward
%NaCl gradients.
%The mean, standard deviation, and skewness for a value of the 
%concentration gradient are calculated from all the runs of a single worm.
%Then, the values from different worms for the same value of the
%concentration gradient are used for calculating their mean and
%standard deviation. The mean values, together with the 
%error bar defined by the one standard deviation range, 
%are plotted in (A)--(C).
%(D) Distributions of $\phi$ conditioned by
%two ranges of lateral NaCl gradient 
%$\mathrm{d}C / \mathrm{d} x_{\mathrm \ell}$. 
%We combined the data of all the worms
%to make the two distributions sufficiently smooth.
}
\label{fig_weathervane_phi}
\end{center}
\end{figure}

In the presence of an NaCl gradient, the run is chiefly characterized by
klinotaxis \cite{Iino2009}, according to which
the worm senses the lateral gradient of NaCl and gradually curves 
toward the direction in which the NaCl concentration increases.
%To generalize our computational model in the case of the environments
%with an NaCl gradient,
We quantified klinotaxis by following the
methods developed in \cite{Iino2009}.  In
Fig.~\ref{fig_weathervane_phi}A, the mean of $\phi$ of single
worms is plotted against the values of the lateral and forward gradients
of NaCl concentration. The error bars in the figure represent the
standard deviations calculated based on the mean values of $\phi$ from
different worms.  Klinotaxis is evident from the
positive correlation between $\phi$ and the lateral NaCl gradient
(i.e., $\mathrm{d}C / \mathrm{d} x_{\mathrm \ell}$).  During the run,
the worm tends to curve its body to encounter a higher concentration of
the NaCl.  Note that $\phi$ and the forward NaCl gradient 
(i.e., $\mathrm{d}C / \mathrm{d} x_{\mathrm f}$) are
almost uncorrelated.

%\begin{figure}
%\begin{center}
%\includegraphics[width=70mm]{fig_weathervane_mean}
%\includegraphics[width=70mm]{fig_weathervane_sigma}
%\includegraphics[width=70mm]{fig_weathervane_skewness}
%\includegraphics[width=70mm]{fig_weathervane_dist}
%\caption{Weathervane mechanism.
%({\it A}) Mean, (B) standard deviation, and (C) skewness 
%of $\phi$ for different values of lateral and forward
%NaCl gradients.
%The mean, standard deviation, and skewness for a value of the 
%concentration gradient are calculated from all the runs of a single worm.
%Then, the values from different worms for the same value of the
%concentration gradient are used for calculating their mean and
%standard deviation. The mean values, together with the 
%error bar defined by the one standard deviation range, 
%are plotted in ({\it A})--(C).
%(D) Distributions of $\phi$ conditioned by
%two ranges of lateral NaCl gradient 
%$\mathrm{d}C / \mathrm{d} x_{\mathrm \ell}$. 
%We combined the data of all the worms
%to make the two distributions sufficiently smooth.}
%\label{fig_weathervane_phi}
%\end{center}
%\end{figure}

Figures~\ref{fig_weathervane_phi}B and C
indicate that
the standard deviation and the skewness of $\phi$ for single worms
are insensitive to
both the lateral and forward NaCl gradients.
The dependence of the skewness on the NaCl gradient
(Fig.~\ref{fig_weathervane_phi}C), which was found previously
\cite{Iino2009}, does exist but is negligible in the present analysis.
Therefore,
we expect that the NaCl gradient effectively shifts the
distribution of $\phi$ by an amount proportional to the lateral NaCl
gradient but hardly changes the shape of the distribution of $\phi$.
Sample distributions of $\phi$ conditioned by two ranges of 
$\mathrm{d}C / \mathrm{d} x_{\mathrm \ell}$ values are compared in
Fig.~\ref{fig_weathervane_phi}D. They do not differ considerably
except for translation, as expected.

\begin{figure}
\begin{center}
\includegraphics[width=70mm]{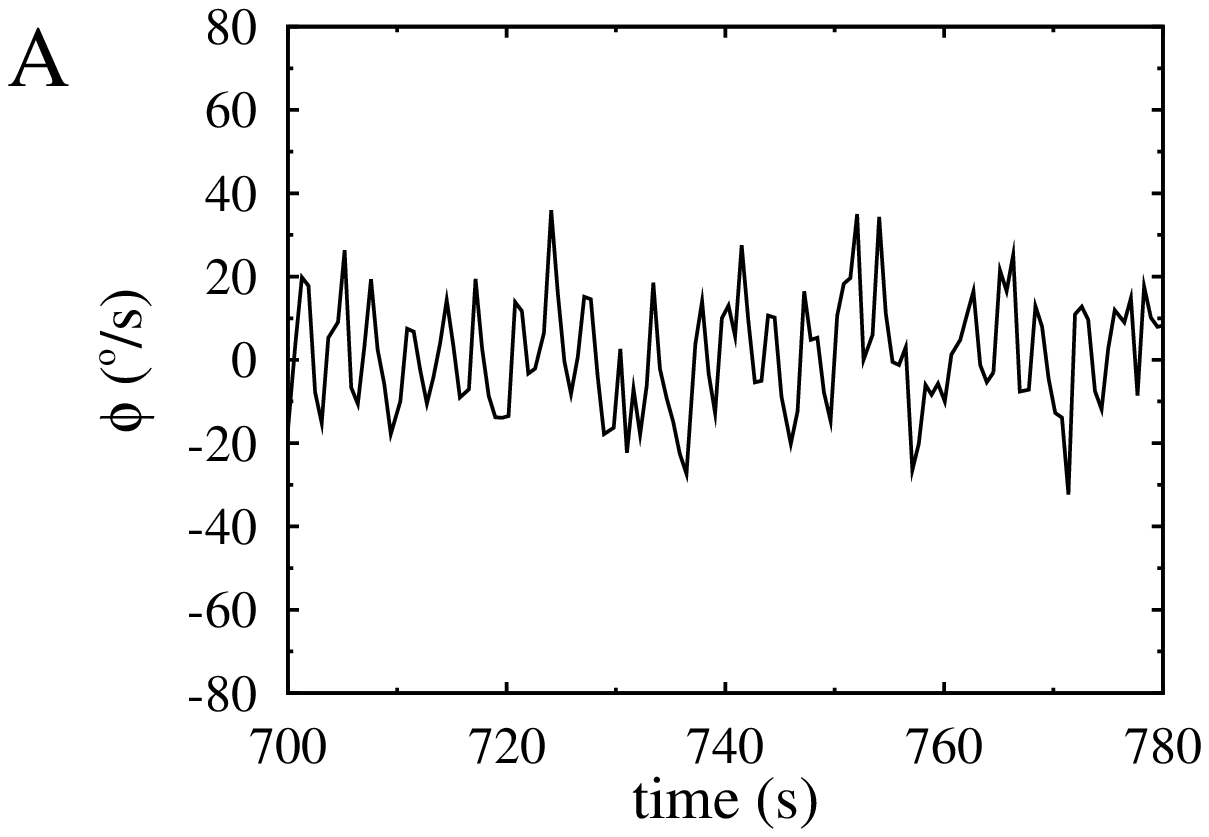}
\includegraphics[width=70mm]{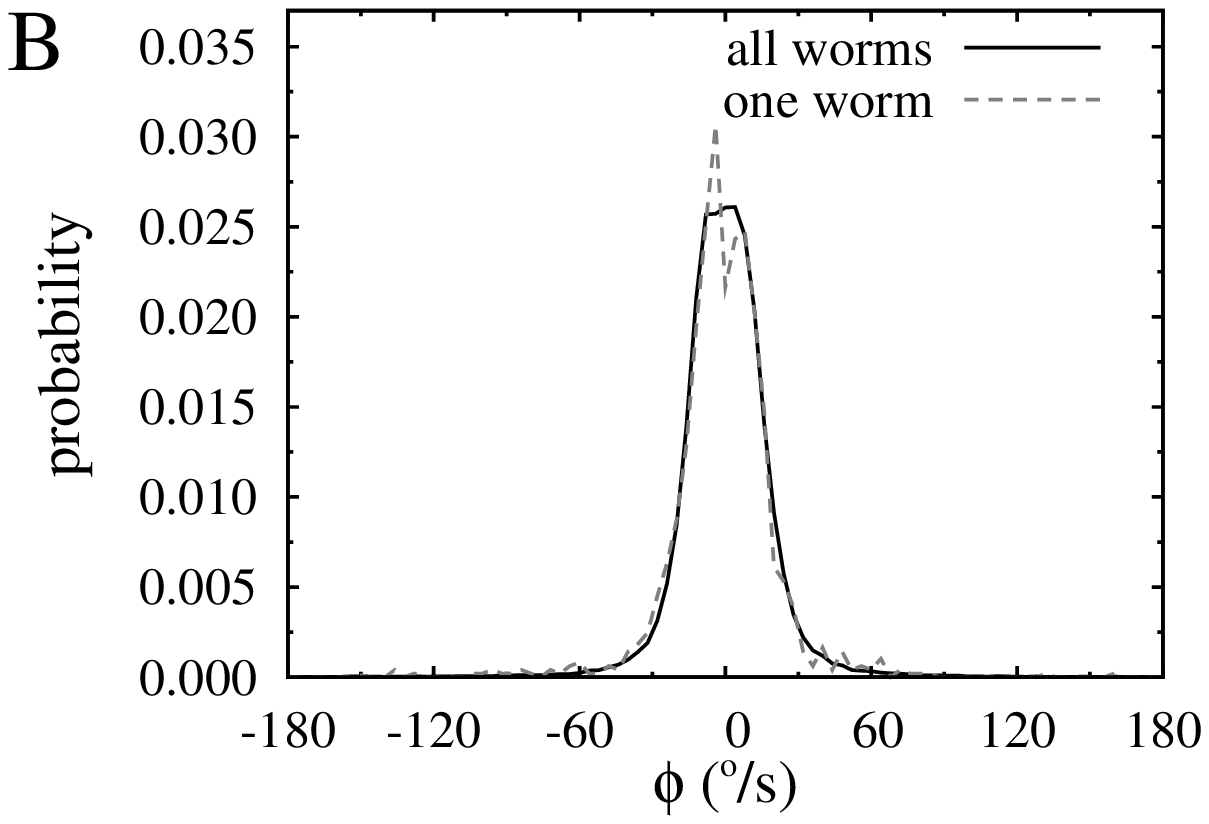}
\includegraphics[width=70mm]{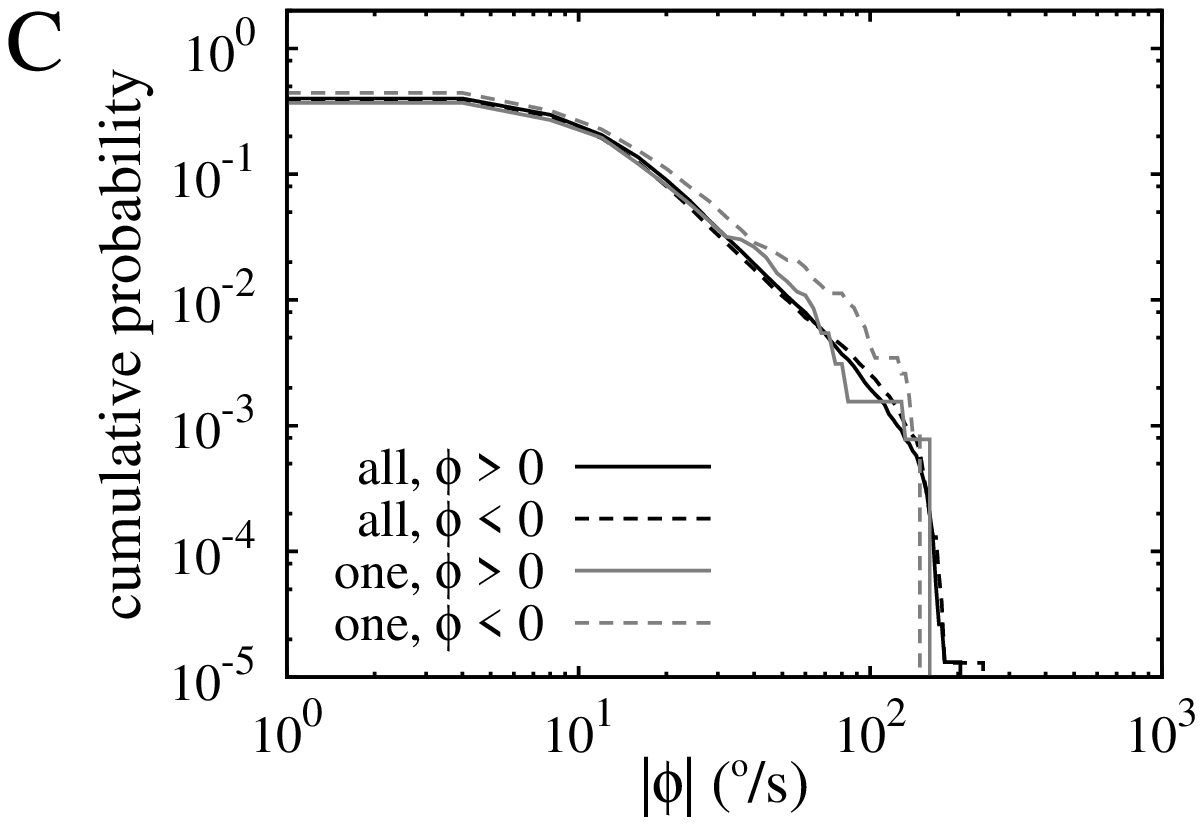}
\caption{
%Results obtained from experiments in the presence of
%an NaCl gradient.
%(A) Sample trajectory of $\phi$.
%(B) Distributions of $\phi$.
%(C) Cumulative distributions of $|\phi|$. See the caption of
%Fig.~\ref{fig phi no-NaCl} for legends.
}
\label{fig phi with NaCl}
\end{center}
\end{figure}

Figure~\ref{fig_weathervane_phi} suggests that
klinotaxis translates the distribution of $\phi$
by a small amount and does not affect the long tails.
A sample trajectory and the distributions of $\phi$ obtained from
the experiments in the presence of an NaCl gradient are shown in
Fig.~\ref{fig phi with NaCl}. The results are qualitatively similar to
those obtained from the experiments
in the absence of an NaCl gradient (Fig.~\ref{fig phi no-NaCl}).

We conclude that the long-tail behavior of the run is robustly observed
both in the absence and presence of an NaCl gradient.

\subsection*{Computational modeling}
  
%\begin{figure}
%\begin{center}
%\includegraphics[width=70mm]{fig_simu_sample_path.eps}
%\includegraphics[width=70mm]{fig_simu_auto_correlation.eps}
%\includegraphics[width=70mm]{fig_simu_distribution.eps}
%\includegraphics[width=70mm]{fig_simu_cumulative_distribution.eps}
%\caption{Results obtained from numerical simulations of the
%computational model in the absence of an NaCl gradient.  
%({\it A}) Sample trajectory of $\phi$ when $T_i$ is constant.  
%(B) Autocorrelation function of $\phi$.
%(C) Distributions of $\phi$.  
%(D) Cumulative distributions of $|\phi|$ on
%the log-log scale.  
%In (B), (C), and (D), the experimental data,
%the numerical results with
%constant $T_i$, and those with fluctuating $T_i$ are
%compared.}
%\label{fig_simu_dist}
%\end{center}
%\end{figure}

%\subsubsection*{\edit{In absence of NaCl gradient: basic model}}

We developed a phenomenological model of the run
on the basis of
the correlated random walk
\cite{Pierce-Shimomura1999,Pierce-Shimomura05,Siniff69,Kitching71,Hall77,Kareiva83,Bovet88jtb,Bartumeus05ecol,Codling08,Iino2009}.

We indexed the discrete time by $i$,
i.e., $t_i = t_{i-1} + \Delta t$, where $\Delta t$ is a time interval.
We modeled the reorientation of the worm at time $t_i$ by
\begin{align}
&\phi_i = R \cos\left(\eta_i \right) 
+ \tilde{\xi}_i,
\label{eq:phi RMP no-NaCl}\\
&\eta_i = \eta_{i-1}+ \frac{2\pi}{T_i} \left(t_i-t_{i-1}\right),
\label{eq:eta RMP no-NaCl}
\end{align}
where $R$, $\eta_i$, and $T_i$ are the amplitude, phase, and 
instantaneous period of undulation, respectively.
$\tilde{\xi}_i$ is the noise inherent at time $t_i$.

If $\tilde{\xi}_i$ were independent Gaussian noise,
as is the case for the standard correlated random walk,
the long-tail behavior of the curving rate is not reproduced.
Therefore, we assumed
multiplicative noise $\tilde{\xi}_i$
\cite{Takayasu97}, such that
\begin{align}
\tilde{\xi}_i = \xi_i \phi_{i-1},
\label{eq_noise_rmp}
\end{align}
where $\xi_i \in \mathcal{N}(0.0, \sigma)$; $\xi_i$ is chosen from the
normal distribution with mean 0 and standard deviation $\sigma$.
The procedure for fitting the parameter values is described in
Appendix I.

The assumption of random multiplicative noise is based on the premise
that a large movement accompanies a proportionally large noise, which
is known for the arm and hand movements of humans
\cite{Sutton67,Schmidt79psyr,Harris98nat,Jones02jnp}.  
To examine the plausibility of this assumption, we consider an
alternative model in which long-tail noise is directly applied in each
time step. If
$\tilde{\xi}_i$ in Eq.~\ref{eq:phi RMP no-NaCl}
is the independent long-tail noise,
the long-tail distribution of $\phi_i$ is produced.
By allowing an autoregressive (AR) term with which to calibrate the autocorrelation of $\phi_i$, we define
the alternative model by replacing
Eq.~\ref{eq:phi RMP no-NaCl} by
$\phi_i = R \cos\left(\eta_i \right) + w \phi_{i-1} + \tilde{\xi}_i$,
where $0 \leq w < 1$.
Note that $\tilde{\xi}_i$ is a long-tail noise;
the AR model with the Gaussian $\tilde{\xi}_i$, as is usually
assumed, does not yield the long-tail distribution
of $\phi_i$ irrespective of the parameter values and the number of
autoregressive terms on the right-hand side of the model.
We compare the two models
%
% in terms of the correlation between the dynamical noise at time $i$ and
%$\phi_{i-1}$ 
%
by measuring
$\left< \left(\phi_i-w\phi_{i-1}\right)^2\right>$,
where $\left< \cdot \right>$ is the average over $i$ for a given
$|\phi_{i-1}|$.
For the alternative long-tail noise model,
we obtain
$\left< \left(\phi_i-w\phi_{i-1}\right)^2\right> =
\left<\left( R \cos\left(\eta_i \right)+\tilde{\xi}_i\right)^2\right>$.
% the sinusoidal
%undulation and the long-tail noise are considered to be
%independent of each other and 
$\tilde{\xi}_i$ 
is identically distributed, and its typical magnitude is much
larger than that of the sinusoidal undulation (i.e.,
$R \cos(\eta_i)$) under the fitted
parameter values. Therefore, $\left<
\left(\phi_i-w\phi_{i-1}\right)^2\right>
\approx \left<\tilde{\xi}_i^2\right>$ would depend little on
$|\phi_{i-1}|$ and serve as an estimate of the magnitude of the noise
for a proper value of $w$.  In contrast, 
for our multiplicative noise model,
\EQS\ref{eq:phi RMP no-NaCl}
and \ref{eq_noise_rmp} imply that $\left<
\left(\phi_i-w\phi_{i-1}\right)^2\right>$ increases with $|\phi_{i-1}|$.

\begin{figure}[h]
\begin{center}
\includegraphics[width=70mm]{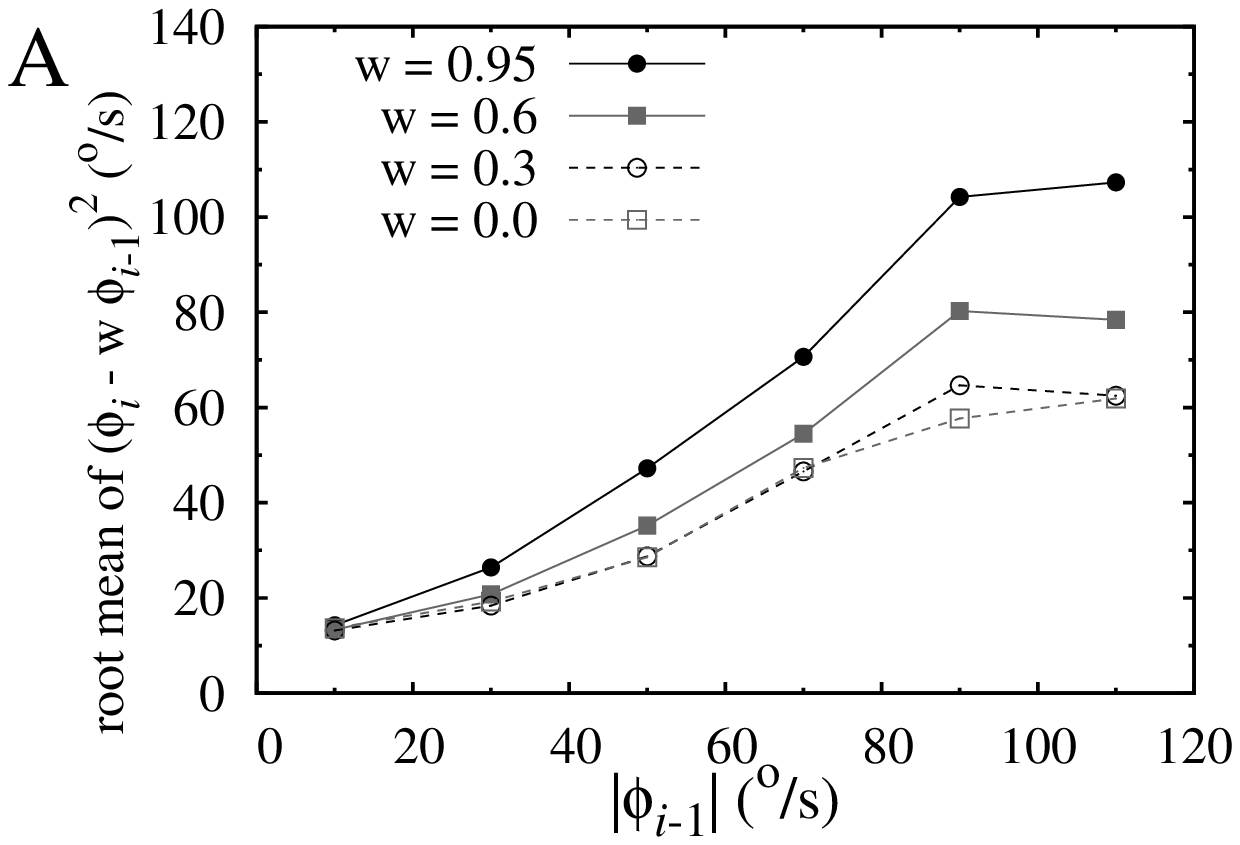}
\includegraphics[width=70mm]{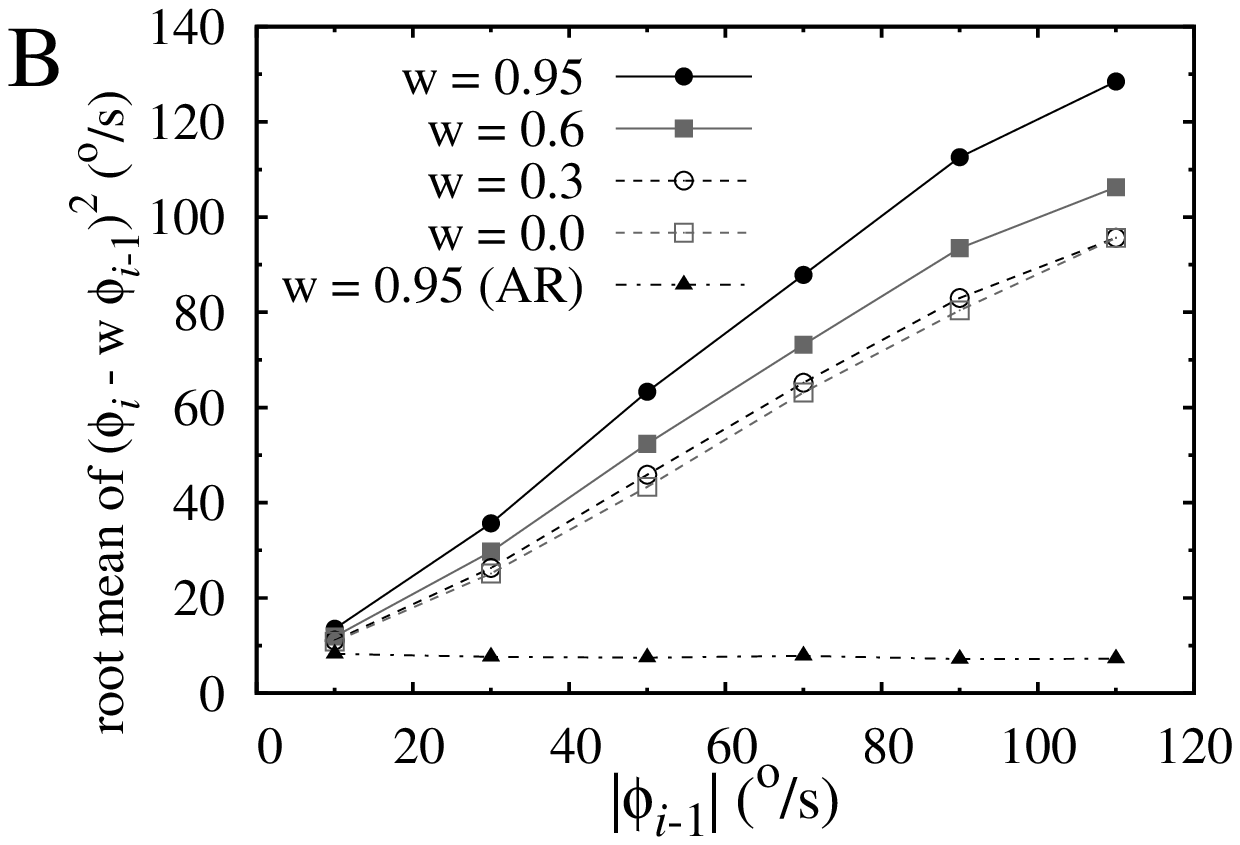}
\caption{}
\label{fig:validate RMP}
\end{center}
\end{figure}

It turns out for any $w$
that $\left<
\left(\phi_i-w\phi_{i-1}\right)^2\right>$ and
$|\phi_{i-1}|$ are positively correlated for both the experimental data
(Fig.~\ref{fig:validate RMP}A) and the multiplicative noise
model (Fig.~\ref{fig:validate RMP}B).  This is not the 
case for the long-tail noise model with $w=0.95$
(Fig.~\ref{fig:validate RMP}B) or any other values of
$w$; the plots
for other values of $w$ almost completely
overlap with those for $w=0.95$ shown in 
Fig.~\ref{fig:validate RMP}B.
Therefore, at least some components of the
locomotion are likely to be explained by the multiplicative noise.
We have verified that the addition of the AR terms to the multiplicative
noise model little affects the behavior of the model
except for some changes in the autocorrelation of $\phi_i$.

\begin{figure}
\begin{center}
\includegraphics[width=70mm]{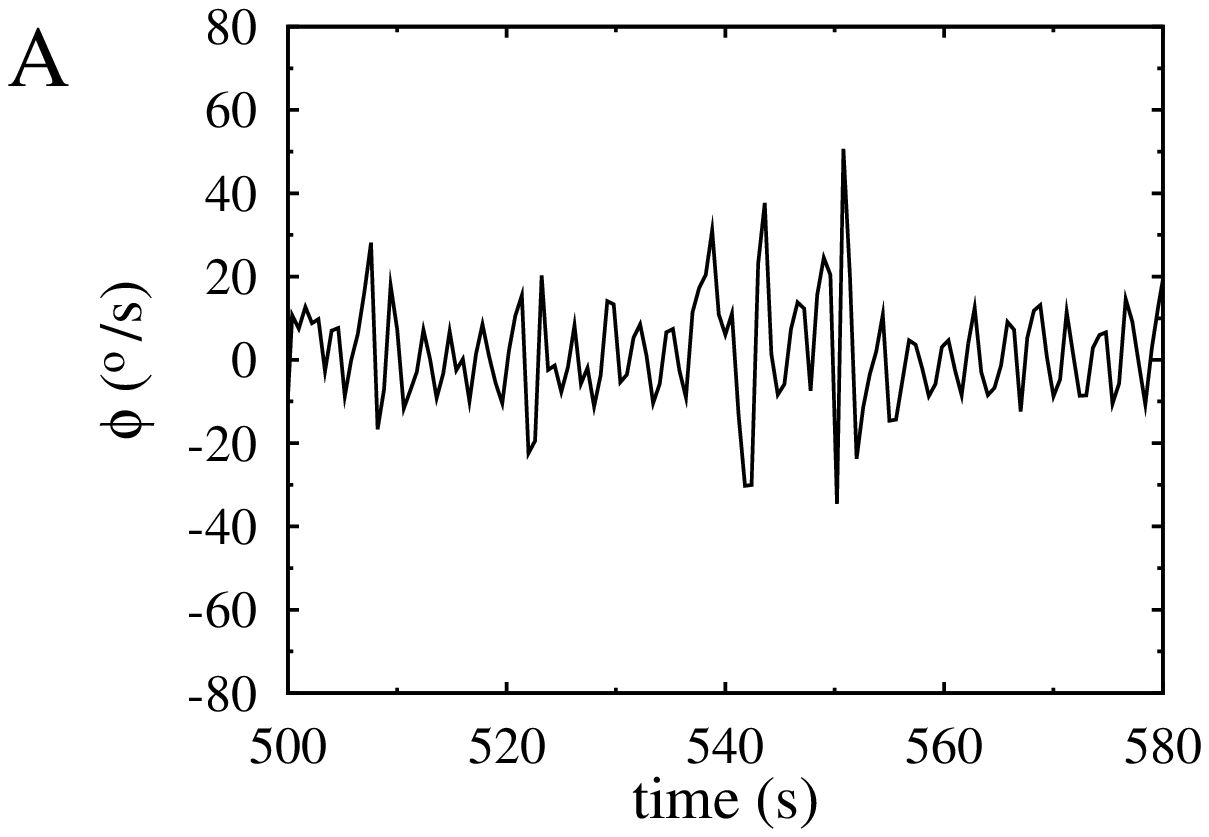}
\includegraphics[width=70mm]{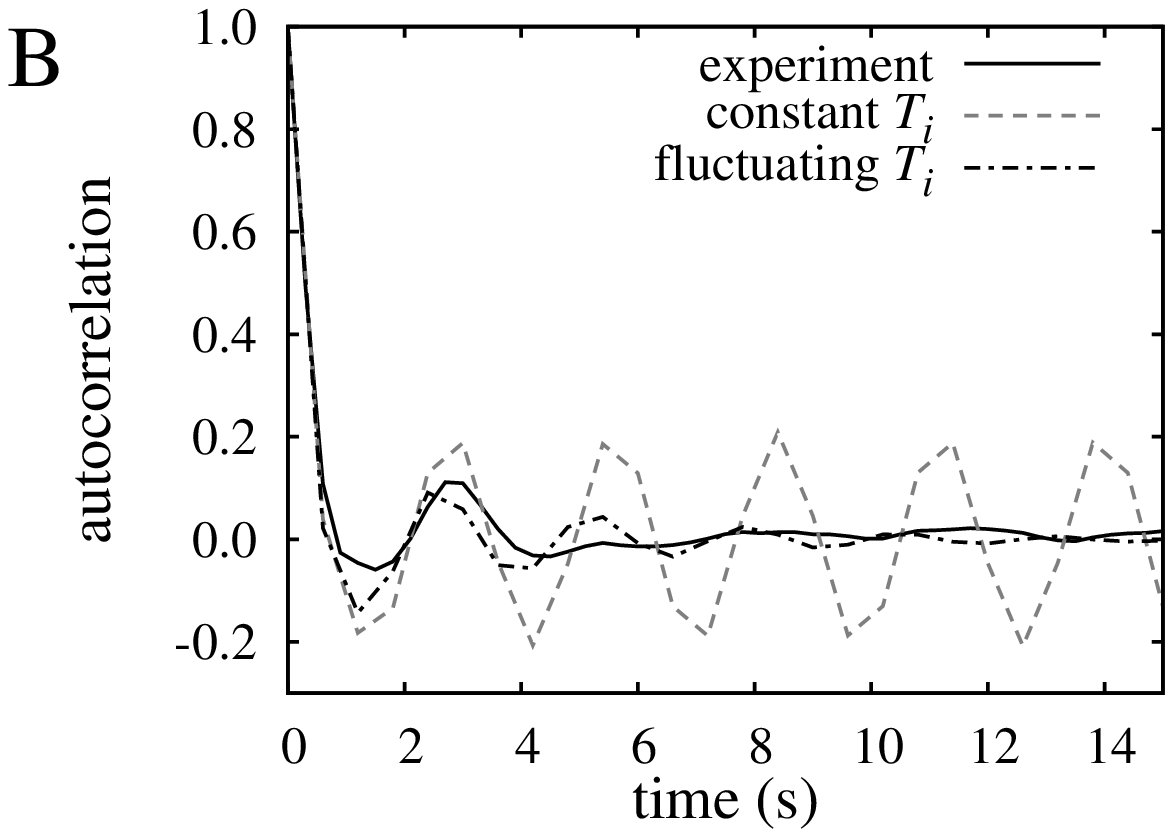}
\includegraphics[width=70mm]{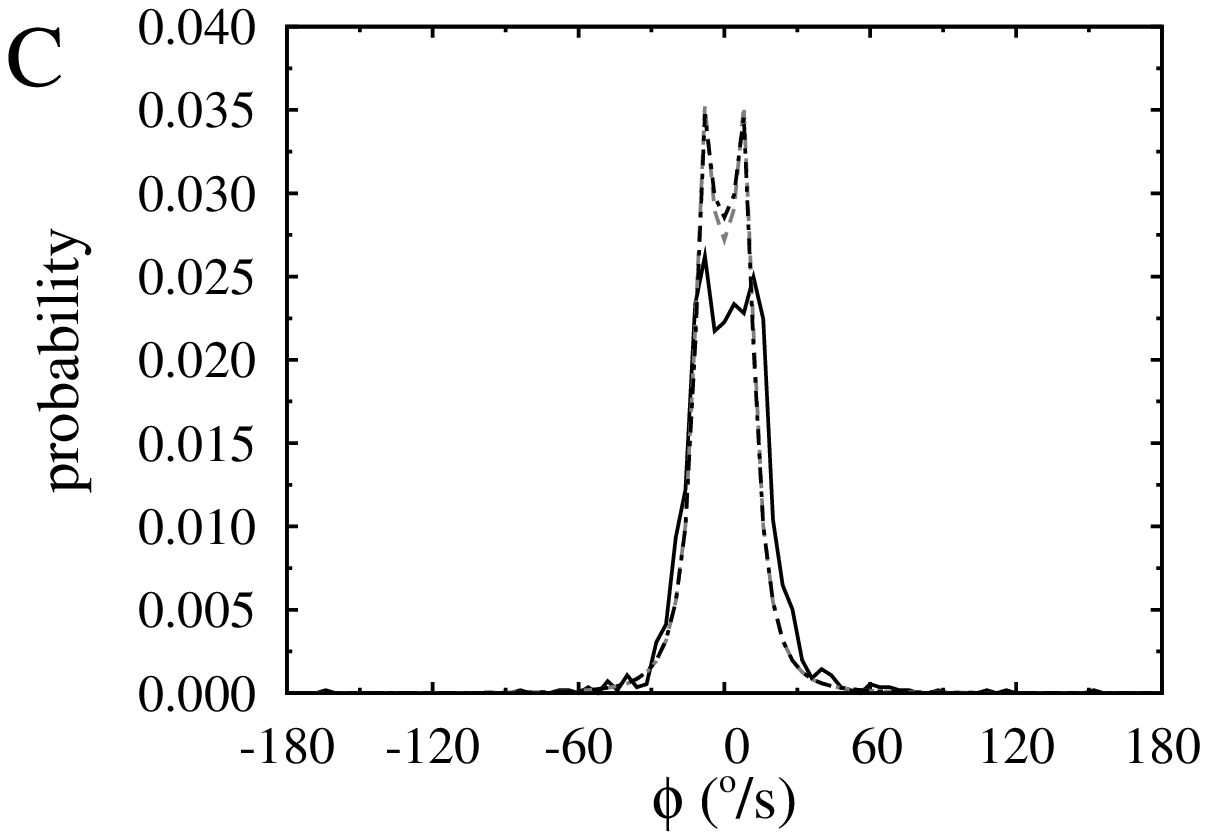}
\includegraphics[width=70mm]{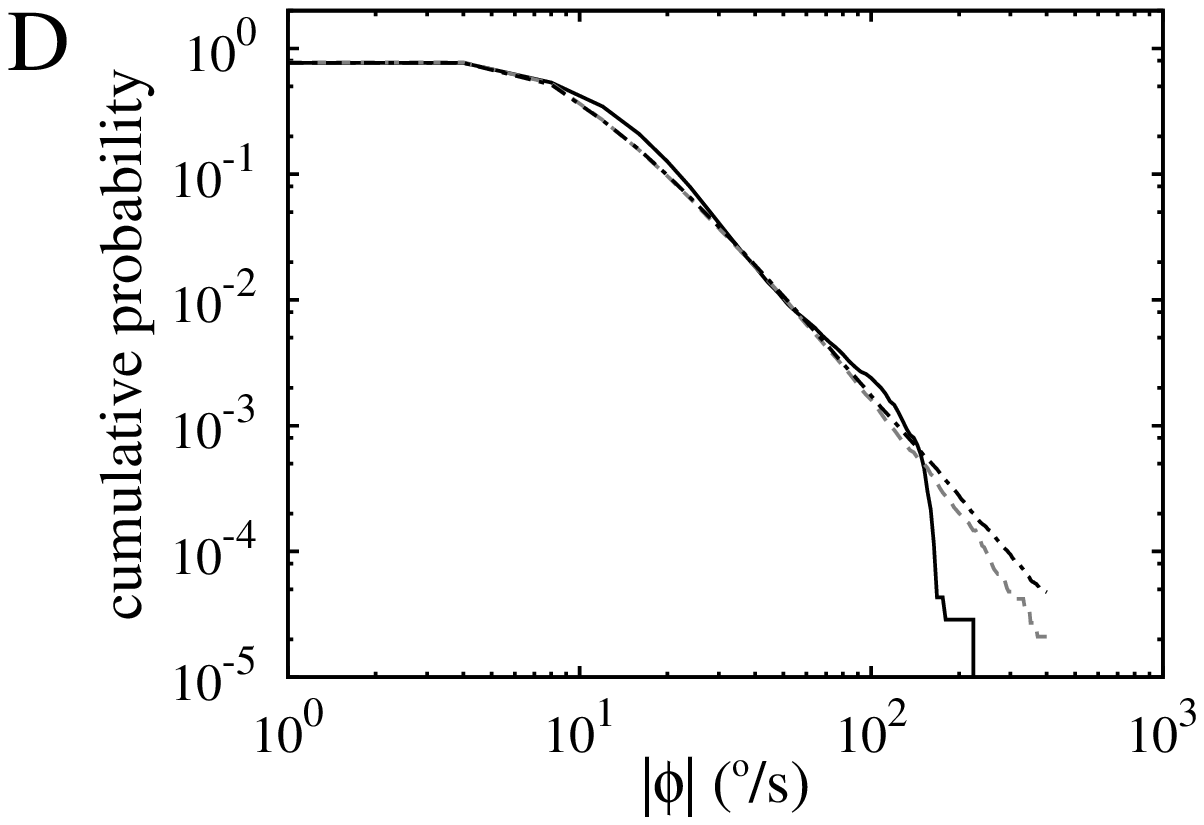}
\caption{
%Results obtained from numerical simulations of the
%computational model in the absence of an NaCl gradient.  
%(A) Sample trajectory of $\phi$ when $T_i$ is constant.  
%(B) Autocorrelation function of $\phi$.
%(C) Distributions of $\phi$.  
%(D) Cumulative distributions of $|\phi|$ on
%the log-log scale.  
%In (B), (C), and (D), the experimental data,
%the numerical results with
%constant $T_i$, and those with fluctuating $T_i$ are
%compared.
}
\label{fig_simu_dist}
\end{center}
\end{figure}

A sample trajectory obtained from the numerical simulations of the
multiplicative noise
model is shown in Fig.~\ref{fig_simu_dist}A, which is
qualitatively similar to that shown in Fig.~\ref{fig phi no-NaCl}A.
In Fig.~\ref{fig_simu_dist}B,
the autocorrelation function obtained from the model
(dashed line labeled ``constant $T_i$'') is compared
with that of the experimental
data (solid line). Even though the model reproduces the
oscillatory nature of the autocorrelation function, the autocorrelation
does not decay, indicating the excessive periodicity inherent in the model.
The strong periodicity results from the fact that
the undulation period $T_i$ is fixed.
The distribution of $\phi$ and its long tail 
obtained from the numerical simulations roughly
resemble those of the experimental data
(Fig.~\ref{fig_simu_dist}C and D).

The undulation period of the experimental data
fluctuates. To incorporate this factor,
we extended the model by allowing $T_i$ to fluctuate with time.
We set $T_i = 2.8 + \zeta$,
where $\zeta \in \mathcal{N}(0.0,0.9)$. 
We generated $T_i$ until a positive value of $T_i$ was obtained.
This modification of the model yielded the exponential
decay in the autocorrelation function, 
as found in the experimental data
(dash-dotted line labeled ``fluctuating $T_i$'' 
in Fig.~\ref{fig_simu_dist}B). 
The modified model 
does not quantitatively agree with the autocorrelation function of
the experimental data, partly because the fluctuation in $T_i$ is crudely implemented.
%we did not explore a further quantitative fit to the autocorrelation
%function for the sake of simplicity.
Note that the modification of the model does not change the
distribution of $\phi$ (dash-dotted lines 
in Figs.~\ref{fig_simu_dist}C and D).  
The sample trajectory is also similar to that
of the original model and that of the experimental data. % (results not shown).  
%
%The random multiplicative noise model with fluctuating period
%appears to be an adequate phenomenological model for the run.

\begin{figure}
\begin{center}
\includegraphics[width=70mm]{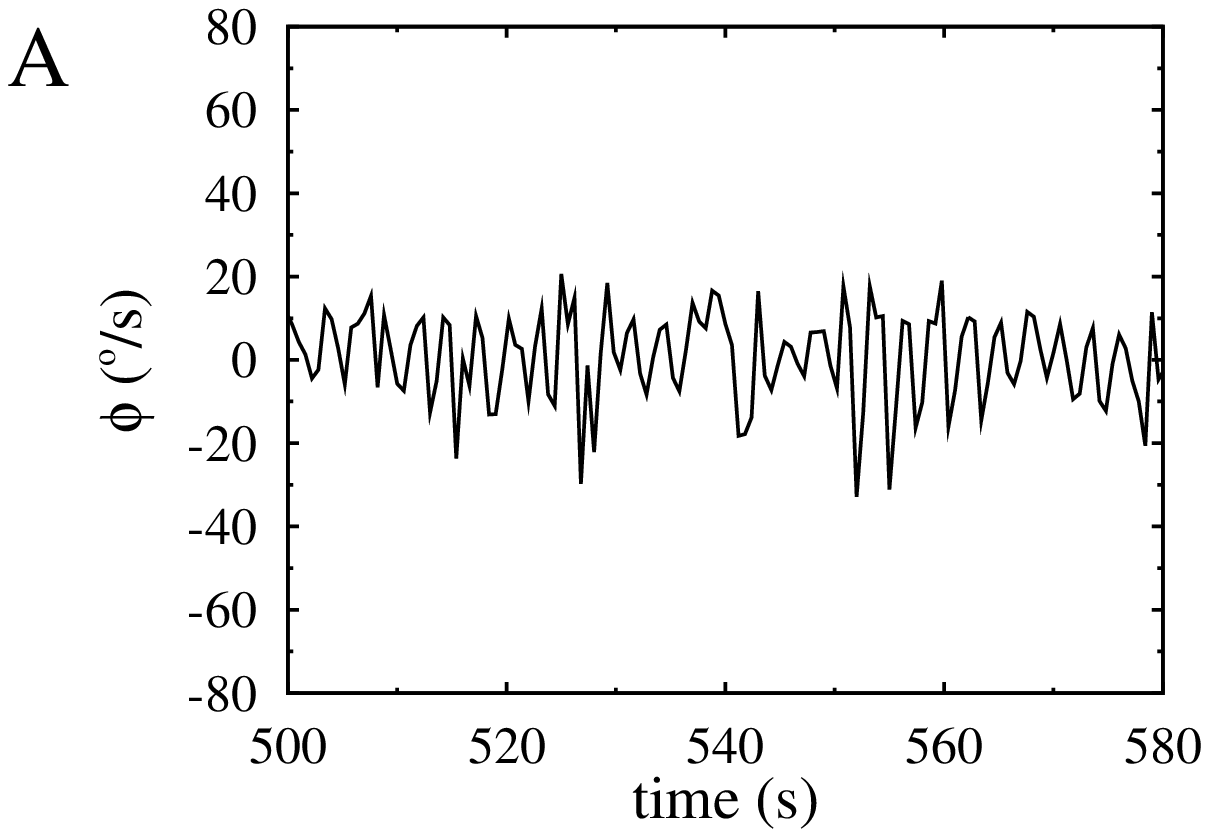}
\includegraphics[width=70mm]{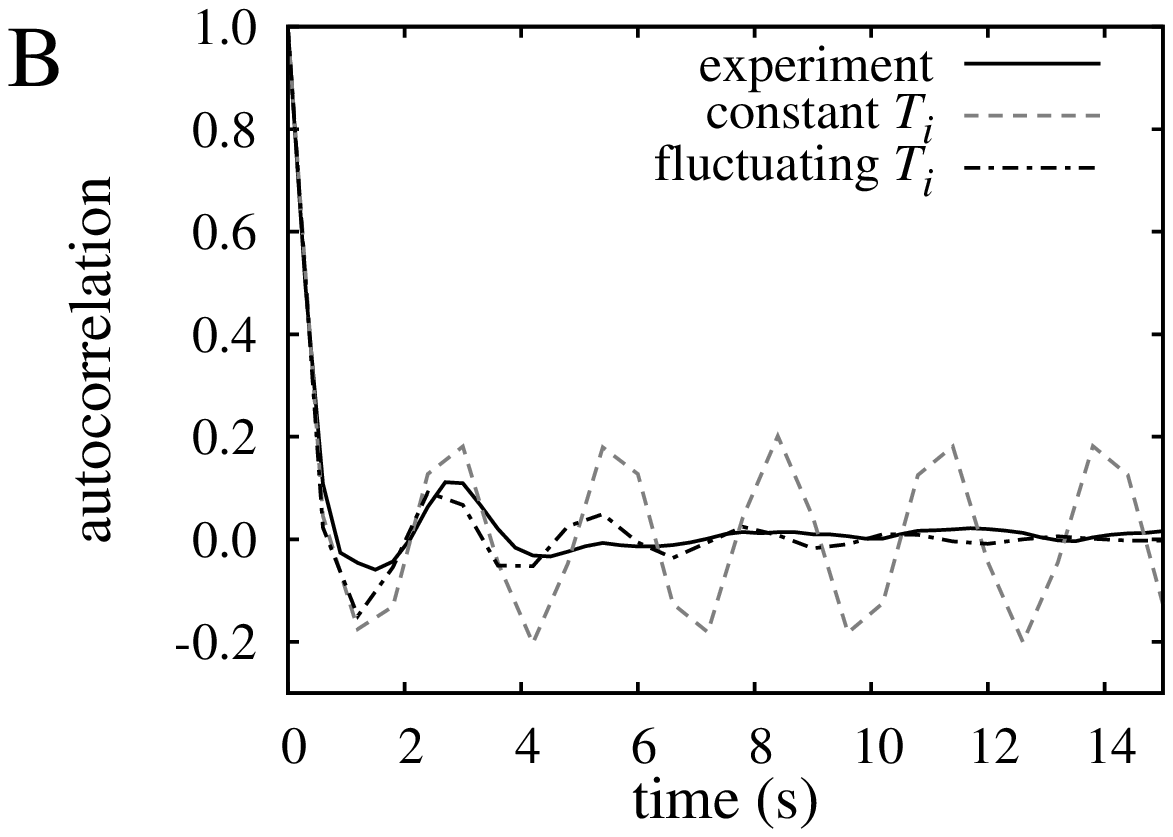}
\includegraphics[width=70mm]{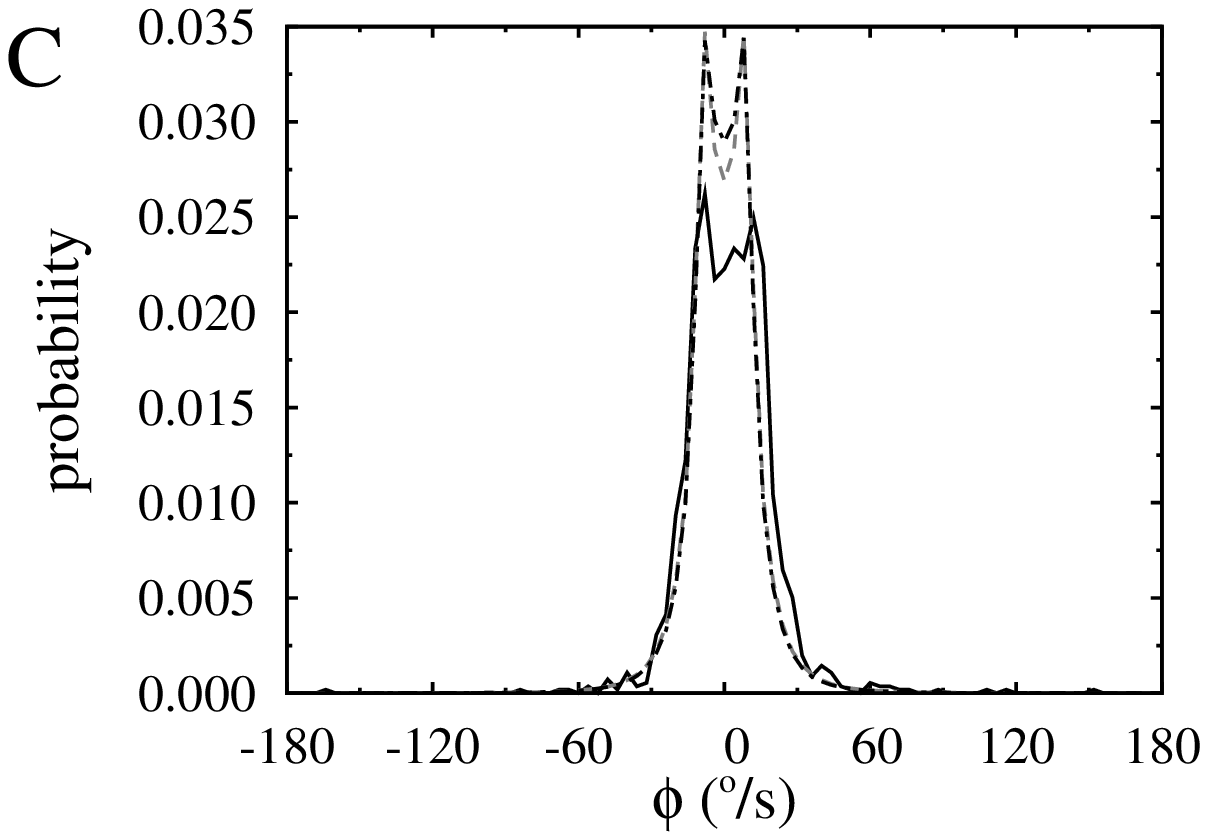}
\includegraphics[width=70mm]{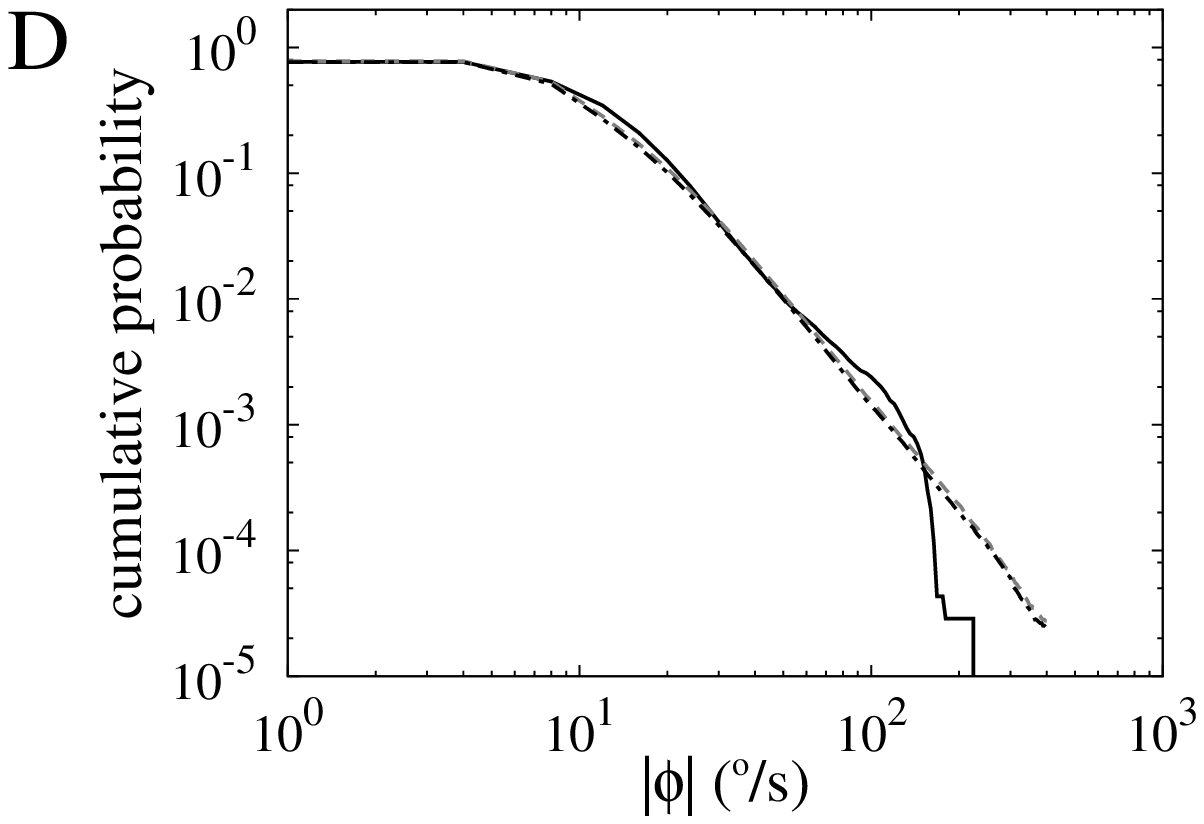}
\caption{
% Numerical results in the absence of an NaCl gradient when
% a different noise model is used. Eq.~\ref{eq_noise_rmp} is
% replaced by
% $\tilde{\xi}_i = \xi_i (\phi^{\prime}_{i-1} + \phi^{\prime}_{i-2}) /2$ 
% with $\xi_i \in \mathcal{N}(0.0,1.24)$. We modified the standard
% deviation of the noise from that of the original model
% (Eq.~\ref{eq_noise_rmp}) to obtain a reasonable fit
% to the experimental data.
% Long-tail distributions of $\phi$ are produced, as shown in
% D. See the caption of 
% Fig.~\ref{fig_simu_dist} for legends.
}
\label{fig:different noise}
\end{center}
\end{figure}

The long-tail behavior is not restricted to the specificity of the
multiplicative noise, 
as defined by Eq.~\ref{eq_noise_rmp}.
The behavior of the model is almost the same
when we replaced Eq.~\ref{eq_noise_rmp}
with a different noise model given by
$\tilde{\xi}_i = \xi_i (\phi_{i-1} + \phi_{i-2}) /2$ 
(Fig.~\ref{fig:different noise}).

%\subsubsection*{\edit{In presence of NaCl gradient}}

%\begin{figure}
%\begin{center}
%\includegraphics[width=70mm]{fig_real_nacl_sample_path.eps}
%\includegraphics[width=70mm]{fig_real_nacl_distribution.eps}
%\includegraphics[width=70mm]{fig_real_nacl_cumulative_distribution.eps}
%\caption{Results obtained from experiments in the presence of
%an NaCl gradient.
%({\it A}) Sample trajectory of $\phi$.
%(B) Distributions of $\phi$.
%(C) Cumulative distributions of $|\phi|$. See the caption of
%Fig.~\ref{fig phi no-NaCl} for legends.}
%\label{fig phi with NaCl}
%\end{center}
%\end{figure}

Under NaCl gradients,
klinotaxis  seems to simply translate
the distribution of $\phi$ (Fig.~\ref{fig_weathervane_phi}).
Therefore, 
we extended the model such that klinotaxis adds linearly to
the random walk. In other words, the curving rate observed under NaCl gradients is
assumed to be the summation of $\phi_i$ given by 
Eq.~\ref{eq:phi RMP no-NaCl} and the klinotaxis term
$\alpha \times \mathrm{d}C/\mathrm{d} x_{\mathrm \ell}$.
%
%We extended
%Eq.~\ref{eq:phi RMP no-NaCl} as follows:
%\begin{align}
%&\phi_i = \phi_i^{\prime}
% + \alpha \frac{\mathrm{d}C}{\mathrm{d} x_{\mathrm \ell}}
%\label{eq_rmp2 with NaCl},
%\end{align}
%%
%where $\alpha$ is the strength of the weathervane mechanism.
%$\phi_i^{\prime}$ is the hidden curving rate at time $i$
%and is equivalent to
%$\phi_i$ defined in Eq.~\ref{eq:phi RMP no-NaCl}, i.e.,
%%
%\begin{align}
%&\phi_i^{\prime} = R \cos\left( \eta_i \right) 
%+ \tilde{\xi}_i,
%\label{eq:phi' RMP with-NaCl}
%\end{align}
%
%where $\eta_i$ is given by Eq.~\ref{eq:eta RMP no-NaCl}.
%The multiplicative noise $\tilde{\xi}_i$ is generated from
%Eq.~\ref{eq_noise_rmp}.
In the numerical simulations, we set 
$\alpha = 19.24$, which is the value of the slope
determined by the linear regression
for the mean of the data shown in Fig.~\ref{fig_weathervane_phi}A.

The numerical results in the presence of an NaCl gradient
almost overlap those from the computational model in the absence of an NaCl gradient
shown in Fig.~\ref{fig_simu_dist}. % (results not shown).
This implies that the model reproduces the long-tail behavior of $\phi$,
as observed in the experimental data in the presence of an NaCl gradient
(Fig.~\ref{fig phi with NaCl}).
We did not calculate the autocorrelation
function because the curving rate is nonstationary
when the worm performs chemotaxis.
We have verified that our model worm performs
chemotaxis as in previous models and experimental data
\cite{Pierce-Shimomura1999,Iino2009}.

\section*{DISCUSSION}

We have analyzed the runs of \textit{C.~elegans}
and discovered
long-tail distributions of the curving rate.
We have proposed a phenomenological model that qualitatively
reproduces the behavior of the curving rate.
Our model is a type of correlated
random walk with multiplicative noise. 

\begin{figure}
\begin{center}
\includegraphics[width=80mm]{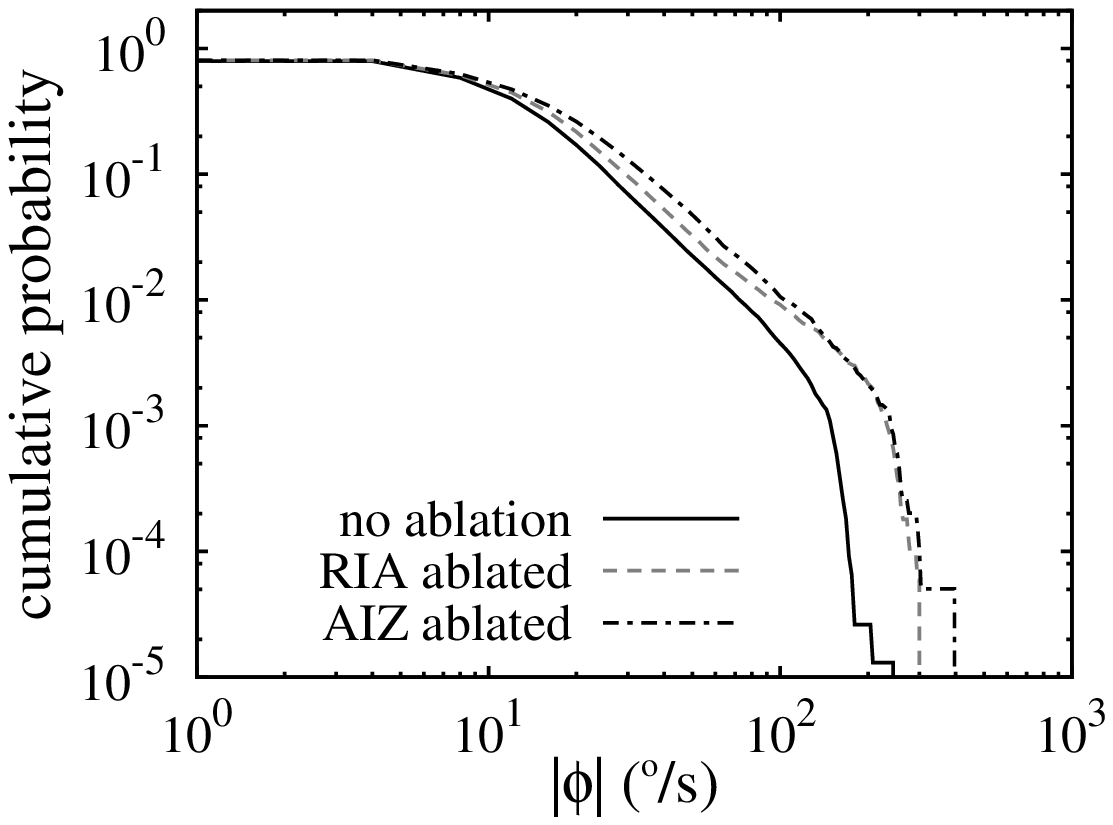}
\caption{
% Cumulative distribution of $|\phi|$ for the worms subjected to
% laser ablation of either of two different neurons (RIA and AIZ). 
% The laser-ablated worms are subjected to an NaCl
% gradient. Each distribution is generated by combining the data of all the worms. The long-tail behavior of the run is supported by these
% data.  The scaling exponent of the long-tail distribution
% depends on the type of ablation.
% The results in the absence of ablation are also plotted for comparison.
}
\label{fig:dist of |phi| ablation}
\end{center}
\end{figure}

The long-tail behavior is robust against various types of
perturbations.
First, it is observed
both in the absence and the presence of NaCl gradients
such that it is independent of klinotaxis.
Second, our results are not restricted to
the specific version of the multiplicative noise
model assumed in this study (see \FIG\ref{fig:different noise}).
Third, the long-tail behavior is robust against
laser ablation of various neurons.
As shown in \FIG\ref{fig:dist of |phi| ablation},
the long-tail behavior is preserved regardless of whether
the ablated neurons are responsible for klinotaxis for NaCl chemotaxis (AIZ neurons; 
see \cite{Iino2009})
or not (RIA neurons). This result gives another support to the observation that the long-tail behavior is independent of klinotaxis.
We speculate that the long-tail behavior reflects intrinsic properties
of the neural circuit for head oscillation, which is currently
unidentified.
Fourth, the bias
in the curving rate that each worm owns
\cite{Pierce-Shimomura1999,Iino2009},
 perhaps because 
the worm crawls on one of its sides,
does not affect the long-tail behavior (see Appendix II).
We have also verified by numerical simulations that
the long-tail behavior of $\phi$ does not hamper klinotaxis.
% of chemotaxis.

The travel length per unit time obeys the L\'{e}vy
distribution (i.e., a power-law distribution) 
for various animals.  Such a L\'{e}vy flight is suggested
to enhance the efficiency of foraging \cite{Viswanathan99,Sim08}.  The
long-tail distributions in the curving rate that we have discovered are
not directly related to the L\'{e}vy flight. In the L\'{e}vy flight,
the distance of the movement per unit time follows a long-tail
distribution. Accordingly, long jumps occasionally occur to possibly accelerate the
foraging. In the locomotion of \textit{C.~elegans}, the distance
of the movement per unit time is narrowly distributed, which reflects
the fact that the worm
moves only by means of regular undulation of the body.  
A long-tail distribution of the curving rate does not appear to
affect the long-term
locomotion behavior relevant in foraging or taxis.

The trajectories of the worms are highly noisy.  The correlated random walk is
a useful tool for modeling noisy migrations of
\textit{C.~elegans}
\cite{Pierce-Shimomura1999,Pierce-Shimomura05,Iino2009} and other
animals \cite{Kitching71,Hall77,Kareiva83,Bovet88jtb,Codling08}.  The
noise per unit time is often modeled as Gaussian noise
\cite{Kitching71,Iino2009}.  However, Gaussian
noise does not lead to distributions with long tails. 
Obviously, we could reproduce the long tail
if independent noise obeying the long-tail distribution
is applied at each time step \cite{Bartumeus05ecol}.
However, this independent-noise model contradicts
the positive correlation between the statistics of the
consecutive curving rates present in the data 
(Fig.~\ref{fig:validate RMP}).
To be consistent with both the long-tail behavior and the
positive correlation, we have
assumed that the amplitude of the noise is proportional to the curving
rate. This assumption is based on the premise that exerting a large force
accompanies a proportionally large noise.  This is the case in the motor
activities of human beings; the amplitude of the physiological noise
is proportional to the magnitude of the hand and arm movements or the
generated force \cite{Sutton67,Schmidt79psyr,Harris98nat,Jones02jnp}.
We note that the positive correlation in the data is not as strong as 
the expectation from a random walk with purely multiplicative noise. 
The locomotion of the actual worm may be generated by combination of
multiplicative noise and independent noise.
To the best of our knowledge, the relationship between the magnitude of the 
force and that of the noise has not
been analyzed for animals other than humans.
An investigation of this issue may be an interesting
future problem. Explicit modeling of the animal's movement
should also be attempted in the future.

%\setcounter{figure}{0} 
%\renewcommand{\thefigure}{A\arabic{figure}}

%\section*{Appendix}

\section*{Appendix I: Determination of parameter values of the
  computational model}

\begin{figure}
\begin{center}
\includegraphics[width=70mm]{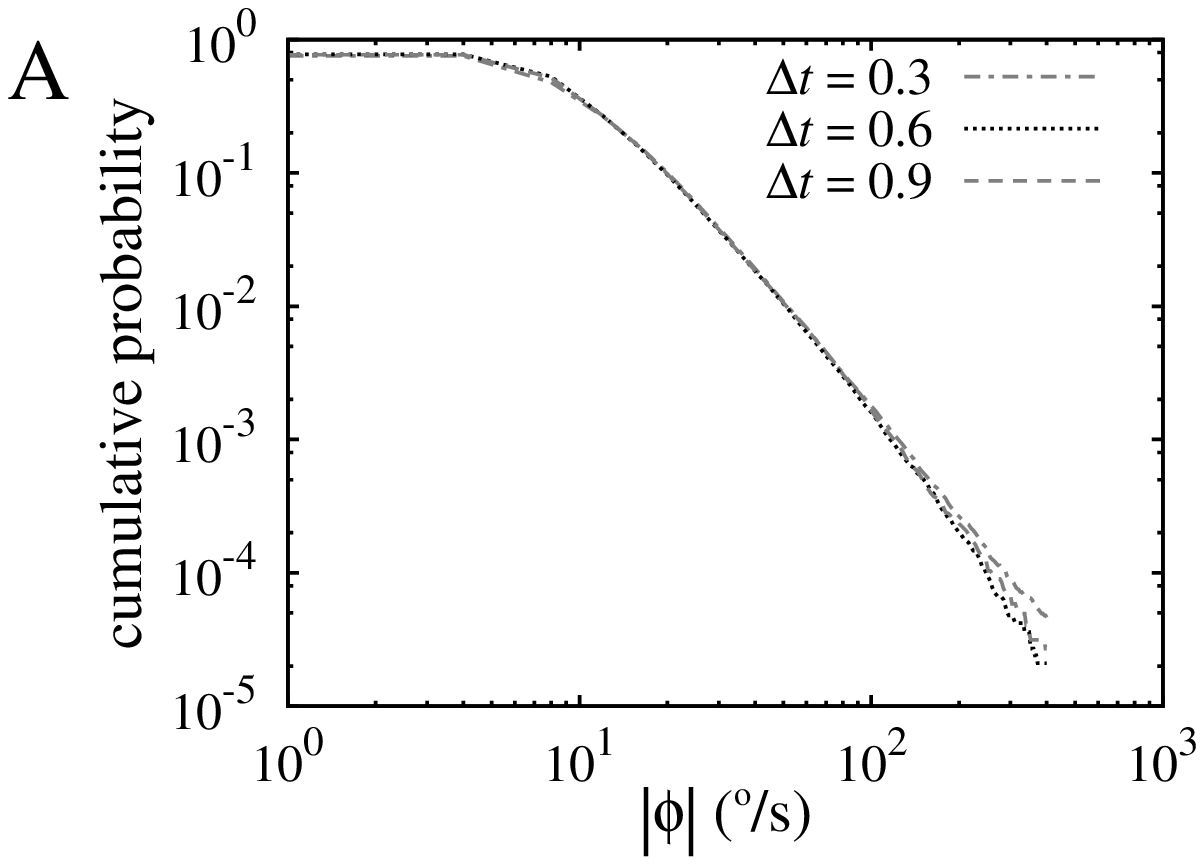}
\includegraphics[width=70mm]{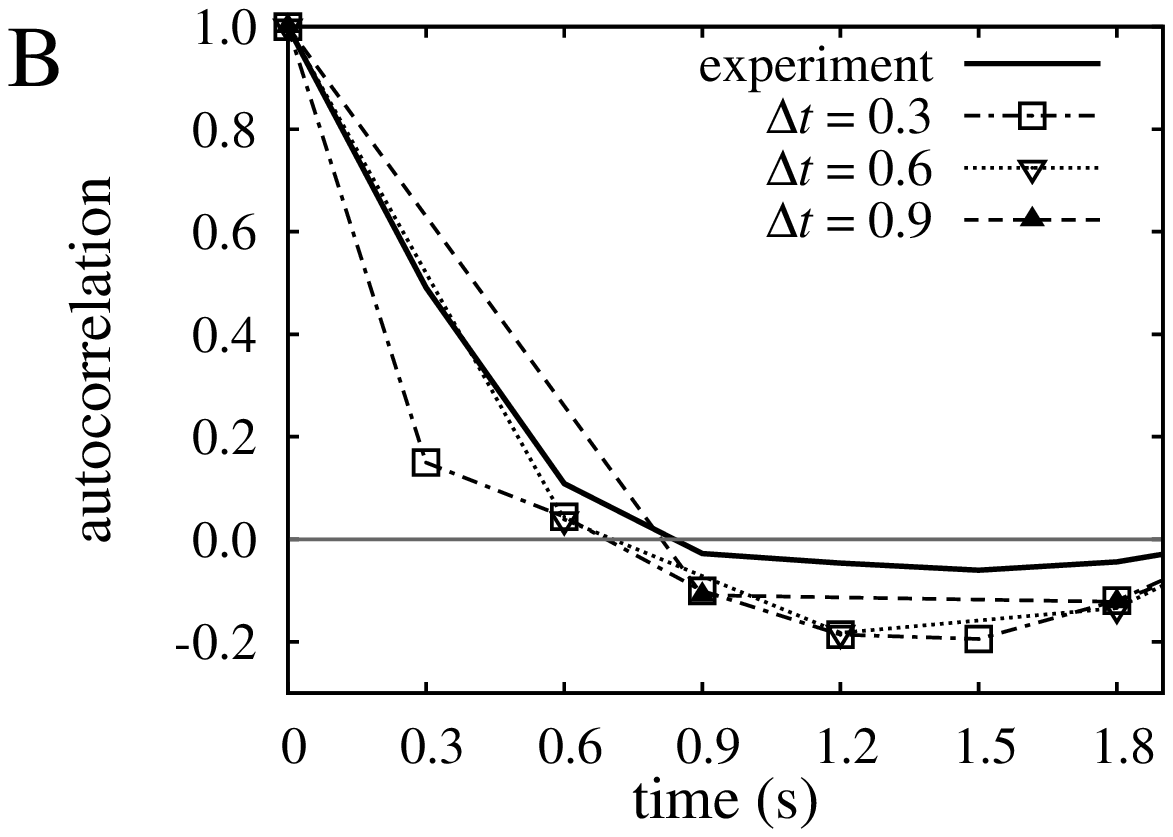}
\caption{
% Numerical results for different values of $\Delta t$.
% (A) Cumulative distributions of $|\phi|$.
% The
% distribution of the curving rate is nearly independent of the value of
% $\Delta t$ in the computational model.
% (B) Autocorrelation function of $\phi$.
% The 
% autocorrelation function for small time lags depends on $\Delta t$.
% A small $\Delta t$ yields a rapid decay of 
% the autocorrelation function for a small lag.
% In the main text, we set $\Delta t=0.6$ s such that the
% autocorrelation function for small time lags is close
% to that of the experimental data.
}
\label{fig_determine_time_interval}
\end{center}
\end{figure}

We determined the four fitting parameters of the model, i.e.,
$\Delta t$, $R$, $T_i$, and $\sigma$, as follows.
On the basis of the long-term oscillatory period of the data,
we set $T_i = 2.8$ s independent of $i$
(also see Fig.~\ref{fig phi no-NaCl}A). 
We also investigated the
case of fluctuating $T_i$ in the main text.
The amplitude of the oscillatory head movement is set to
$R = 10.0$\degs so that the dual peaks
in Fig.~\ref{fig phi no-NaCl}C are roughly recovered.
These two parameters are necessary for reproducing
the undulation of the worm.
The standard deviation of the intrinsic noise (i.e., $\sigma$)
controls the 
dispersion of $\phi$. 
In order to match the standard deviation of the distribution of $\phi$
with the experimental one, we set $\sigma= 0.9$.
The parameters $R$ and $\sigma$ mainly control
the distribution of $\phi$ for small $|\phi|$ and large $|\phi|$,
 respectively.
It was found that $\sigma$ does not affect the timescale of the
autocorrelation to a great extent.
$\Delta t$ is a relevant parameter of the present model and
mainly affects the timescale of the decay
of the autocorrelation function of $\phi$ for small time lags
(see Fig.~\ref{fig_determine_time_interval}). To fit the experimental data in this aspect,
we set $\Delta t = 0.6$ s.

\section*{Appendix II: Worm-dependent bias in curving rate}
%\subsection*{Appendix IV: Worm-dependent bias in the curving rate}

\begin{figure}
\begin{center}
\includegraphics[width=70mm]{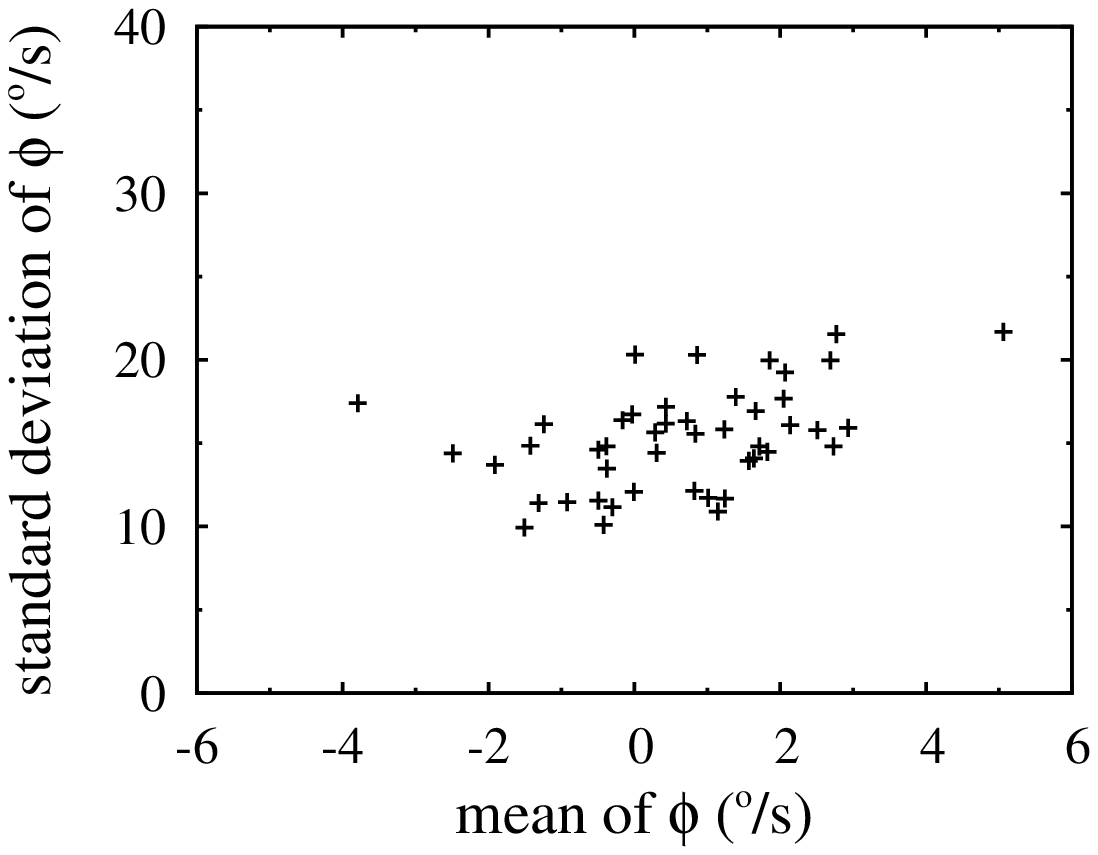}
\caption{
% Relationship between the 
% standard deviation of $\phi$ and
% the mean of $\phi$ for each worm in the absence of an NaCl gradient.
}
\label{fig_real_mean_sigma}
\end{center}
\end{figure}

The standard deviation of 
$\phi$ in the absence of an NaCl gradient
is plotted against the mean of $\phi$
in Fig.~\ref{fig_real_mean_sigma}. One data point corresponds to
one worm.
Figure~\ref{fig_real_mean_sigma} indicates that
the bias (i.e., mean $\phi$) inherent in each worm
is small relative to its standard deviation. However, the
effect of this bias is sufficiently large to generate a visible rotating
trend of the locomotion. % (results not shown).

To examine the effect of the bias on our main
results, we considered the curving rate generated by
adding a constant bias as well as the klinotaxis term
$\alpha \times \mathrm{d}C/\mathrm{d} x_{\mathrm \ell}$ to
the right-hand side of Eq.~\ref{eq:phi RMP no-NaCl}
\cite{Pierce-Shimomura1999,Pierce-Shimomura05}.
We confirmed that this modification of the model
does not significantly affect the distribution of $\phi$ and 
the shape of the autocorrelation function, regardless of
the magnitude of the constant bias. % (results not shown).
The bias has a significant effect on a long time scale, but not on a short one.
Because our analysis of $\phi$ is concerned with 
a short time scale, we did not detrend the bias in the analysis of
our experimental data in the main text.

Note that the concept of the bias discussed here
is different from the tendency that the worm
migrates in a specific direction in which the concentration of the
attractant increases \cite{Codling08}.

\section*{Acknowledgments}

J.O., N.M., and Y.I.
acknowledge the support through
the Grants-in-Aid for Scientific Research on Innovative Areas ``Systems Molecular Ethology''
(Nos. 20115002 and 20115009) from MEXT, Japan.

\newpage
\clearpage

\section*{FIGURE LEGENDS}

\noindent
Figure 1:
(A) Sample trajectory of the worm.
(B) Definition of the curving rate $\phi_i$.
(C) Definition of the sharp turn segment (shaded area).

\vspace{10mm}

\noindent
Figure 2:
Results obtained from experiments in the absence of
an NaCl gradient.
(A) Sample trajectory of $\phi$.
(B) Autocorrelation function of $\phi$.
(C) Distributions of $\phi$
based on the runs of a single worm and those of all the worms.
We use a bin width of 4\degs.
(D) Cumulative distributions of $|\phi|$ (i.e., fraction of values above $|\phi|$)
on the log-log scale.
The results for one worm and the combined one for all the worms are shown
separately for positive and negative values of $\phi$.
The dot-dashed line in (D) indicates
the cumulative distribution of the Gaussian distribution with mean
zero and standard deviation 15.8. 
(E) Distribution of $|\phi|$ for the data
before and after sharp turns are excluded. The data for all the worms are
combined to generate each distribution.

\vspace{10mm}

\noindent
Figure 3:
(A) Relationship between the 
mean of $|\phi|$ and the
instantaneous speed of the worms. The mean values are
calculated from all the runs of a single worm
by categorizing the data into bins of width 0.025 mm/s
and taking the average in each bin.
Then, the mean values from different worms for the same bin
are used for calculating the grand mean and
standard deviation. The grand mean and
error bar defined by the one standard deviation range 
are plotted in (A).
(B) Cumulative distribution of $|\phi|$ when
data points for which the speed of the worms is less than 0.05
mm/s are excluded.
(C) Cumulative distributions of $|\phi|$ when the
data points within 6 s before or after each sharp turn are excluded.
For each figure, the data obtained
from all the worms are combined to generate the cumulative
distribution.

\vspace{10mm}

\noindent
Figure 4:
Klinotaxis.
(A) Mean, (B) standard deviation, and (C) skewness 
of $\phi$ for different values of lateral and forward
NaCl gradients.
The mean, standard deviation, and skewness for a value of the 
concentration gradient are calculated from all the runs of a single worm.
Then, the values from different worms for the same value of the
concentration gradient are used for calculating their mean and
standard deviation. The mean values, together with the 
error bar defined by the one standard deviation range, 
are plotted in (A)--(C).
(D) Distributions of $\phi$ conditioned by
two ranges of lateral NaCl gradient 
$\mathrm{d}C / \mathrm{d} x_{\mathrm \ell}$. 
We combined the data of all the worms
to make the two distributions sufficiently smooth.

\vspace{10mm}

\noindent
Figure 5:
Results obtained from experiments in the presence of
an NaCl gradient.
(A) Sample trajectory of $\phi$.
(B) Distributions of $\phi$.
(C) Cumulative distributions of $|\phi|$. See the caption of
Fig.~\ref{fig phi no-NaCl} for legends.

\vspace{10mm}

\noindent
Figure 6:
%Numerical results for different values of $\Delta t$.
%(A) Cumulative distributions of $|\phi|$.
%(B) Autocorrelation function of $\phi$.
Testing the multiplicative noise hypothesis.
We plot $\sqrt{\left<\left(\phi_i - w \phi_{i-1}\right)^2\right>}$
against $|\phi_{i-1}|$ for four values of $w$ for (A) experimental
data and (B) random walk model with multiplicative noise.
The results for the AR model with $w=0.95$ and the power-law noise
distribution with exponent 3 are also shown in (B).
Each plotted point represents 
the square root of the average of $\left(\phi_i - w \phi_{i-1}\right)^2$ in a bin. Each bin has width 20\degs and contains
the corresponding values of $|\phi_{i-1}|$ obtained from all the worms.

\vspace{10mm}

\noindent
Figure 7:
Results obtained from numerical simulations of the
computational model in the absence of an NaCl gradient.  
(A) Sample trajectory of $\phi$ when $T_i$ is constant.  
(B) Autocorrelation function of $\phi$.
(C) Distributions of $\phi$.  
(D) Cumulative distributions of $|\phi|$ on
the log-log scale.  
In (B), (C), and (D), the experimental data,
the numerical results with
constant $T_i$, and those with fluctuating $T_i$ are
compared.

\vspace{10mm}

\noindent
Figure 8:
Numerical results in the absence of an NaCl gradient when
a different noise model is used. Eq.~\ref{eq_noise_rmp} is
replaced by
$\tilde{\xi}_i = \xi_i (\phi_{i-1} + \phi_{i-2}) /2$ 
with $\xi_i \in \mathcal{N}(0.0,1.24)$. We modified the standard
deviation of the noise from that of the original model
(Eq.~\ref{eq_noise_rmp}) to obtain a reasonable fit
to the experimental data.
Long-tail distributions of $\phi$ are produced, as shown in
D. See the caption of 
Fig.~\ref{fig_simu_dist} for legends.

\vspace{10mm}

\noindent
Figure 9:
Cumulative distribution of $|\phi|$ for the worms subjected to
laser ablation of either of two different neurons (RIA and AIZ). 
The laser-ablated worms are subjected to an NaCl
gradient. Each distribution is generated by combining the data of all the worms. The long-tail behavior of the run is supported by these
data.  The scaling exponent of the long-tail distribution
depends on the type of ablation.
The results in the absence of ablation are also plotted for comparison.

\vspace{10mm}

\noindent
Figure 10:
Numerical results for different values of $\Delta t$.
(A) Cumulative distributions of $|\phi|$.
The
distribution of the curving rate is nearly independent of the value of
$\Delta t$ in the computational model.
(B) Autocorrelation function of $\phi$.
The 
autocorrelation function for small time lags depends on $\Delta t$.
A small $\Delta t$ yields a rapid decay of 
the autocorrelation function for a small lag.
In the main text, we set $\Delta t=0.6$ s such that the
autocorrelation function for small time lags is close
to that of the experimental data.

\vspace{10mm}

\noindent
Figure 11:
Relationship between the 
standard deviation of $\phi$ and
the mean of $\phi$ for each worm in the absence of an NaCl gradient.

\newpage

\section*{FIGURES}

\vspace{10mm}

\newpage
\clearpage

\newpage
\clearpage

\newpage
\clearpage

\newpage
\clearpage

\newpage
\clearpage

\newpage
\clearpage

\newpage
\clearpage

\newpage
\clearpage

\newpage
\clearpage

\newpage
\clearpage

\end{document}